%% file: main.tex
\documentclass[fleqn]{aa}  

\usepackage{graphicx}
\usepackage{xcolor}
\usepackage{hyperref}
\usepackage{amsmath}
\usepackage{txfonts}
\usepackage{calligra}
\usepackage{calrsfs}
\usepackage[mathcal]{eucal}
\usepackage[T1]{fontenc}
\usepackage{longtable}
\usepackage{enumitem}

\newcommand\micron{\mbox{$\mu$m}}
\newcommand{\Msun}{${\rm M}_{\odot}$\,}
\newcommand{\Lsun}{${\rm L}_{\odot}$}

\newcommand{\K}{\, {\rm K}}
\newcommand{\mmas}{\hbox{$\mu$as}}

\begin{document} 

   \title{JWST's PEARLS: Mothra, a new kaiju star at $z=2.091$  extremely magnified by MACS0416, and implications for dark matter models.}

   \titlerunning{Mothra, a new kaiju star}
   \authorrunning{Diego et al.}
 
   
   \author{J.M. Diego\inst{1}\fnmsep\thanks{jdiego@ifca.unican.es}
    \and
Bangzheng Sun\inst{2}
    \and
Haojing Yan\inst{2}
    \and
Lukas J. Furtak\inst{3}
    \and
Erik Zackrisson\inst{4}
    \and
 Liang Dai\inst{5}
  \and
Patrick Kelly\inst{6}
    \and
Mario Nonino\inst{7}
    \and
Nathan Adams\inst{8}
    \and
Ashish K. Meena\inst{3}
    \and
S. P. Willner\inst{9}
    \and
Adi Zitrin\inst{3}
    \and
Seth H. Cohen\inst{10} 
    \and
Jordan C. J. D'Silva\inst{11,12} 
    \and
Rolf A. Jansen\inst{10} 
    \and
Jake Summers\inst{10} 
    \and
Rogier A. Windhorst\inst{10} 
    \and
Dan Coe\inst{13,14,15} 
    \and
Christopher J. Conselice\inst{8} 
    \and
Simon P. Driver\inst{14} 
    \and
Brenda Frye\inst{16} 
    \and
Norman A. Grogin\inst{10} 
    \and
Anton M. Koekemoer\inst{10} 
    \and
Madeline A. Marshall\inst{17,15} 
    \and
Mario Nonino\inst{18} 
    \and
Nor Pirzkal\inst{10} 
    \and
Aaron Robotham\inst{14} 
    \and
Michael J. Rutkowski\inst{19} 
    \and
Russell E. Ryan, Jr.\inst{10} 
    \and
Scott Tompkins\inst{13} 
    \and
Christopher N. A. Willmer\inst{16} 
  \and
  Rachana Bhatawdekar\inst{20}
 }   
\institute{Instituto de F\'isica de Cantabria (CSIC-UC). Avda. Los Castros s/n. 39005 Santander, Spain
\and
Department of Physics and Astronomy, University of Missouri, Columbia, MO 65211, USA 
   \and
Physics Department, Ben-Gurion University of the Negev, P.O. Box 653, Be’er-Sheva 84105, Israel 
   \and
Observational Astrophysics, Department of Physics and Astronomy, Uppsala University, Box 516, SE-751 20 Uppsala, Sweden 
   \and
Department of Physics, 366 Physics North MC 7300, University of California, Berkeley, CA 94720, USA 
   \and
 Minnesota Institute for Astrophysics, University of Minnesota, 116 Church Street SE, Minneapolis, MN 55455, USA 
   \and
INAF-Trieste Astronomical Observatory, Via Bazzoni 2, IT-34124, Trieste, Italy 
   \and
Jodrell Bank Centre for Astrophysics, Alan Turing Building, University of Manchester, Oxford Road, Manchester M13 9PL, UK 
   \and
Center for Astrophysics \textbar\ Harvard \& Smithsonian, 60 Garden St., Cambridge, MA 02138 USA 
   \and
School of Earth and Space Exploration, Arizona State University, Tempe, AZ 85287-1404, USA 
\and
International Centre for Radio Astronomy Research (ICRAR) and the International Space Centre (ISC), The University of Western Australia, M468, 35 Stirling Highway, Crawley, WA 6009, Australia 
\and
ARC Centre of Excellence for All Sky Astrophysics in 3 Dimensions (ASTRO 3D), Australia 
\and
   Space Telescope Science Institute, 3700 San Martin Drive, Baltimore, MD 21218, USA 
\and
Association of Universities for Research in Astronomy (AURA) for the European Space Agency (ESA), STScI, Baltimore, MD 21218, USA 
\and
Center for Astrophysical Sciences, Department of Physics and Astronomy, The Johns Hopkins University, 3400 N Charles St. Baltimore, MD 21218, USA 
\and
Steward Observatory, University of Arizona, 933 N Cherry Ave, Tucson, AZ, 85721-0009, USA 
\and
National Research Council of Canada, Herzberg Astronomy \& Astrophysics Research Centre, 5071 West Saanich Road, Victoria, BC V9E 2E7, Canada 
\and
INAF-Osservatorio Astronomico di Trieste, Via Bazzoni 2, 34124 Trieste, Italy 
\and
Minnesota State University-Mankato,  Telescope Science Institute, TN141, Mankato MN 56001, USA 
\and
European Space Agency (ESA), European Space Astronomy Centre (ESAC), Camino Bajo del Castillo s/n, 28692 Villanueva de la Cañada, Madrid, Spain
}

 \abstract{
   We report the discovery of Mothra, an extremely magnified monster star, likely a binary system of two supergiant stars, in one of the strongly lensed galaxies behind the galaxy cluster MACS0416.  The star is in a galaxy with spectroscopic redshift $z=2.091$ in a portion of the galaxy that is parsecs away from the cluster caustic. The binary star is observed only on the side of the critical curve with negative parity but has been detectable for at least eight years, implying the presence of a small lensing perturber. 
  Microlenses alone cannot explain the earlier observations of this object made with the Hubble Space Telescope. A larger perturber with a mass of at least $10^4$\,\Msun\ offers a more satisfactory explanation. Based on the lack of perturbation on other nearby sources in the same arc, the maximum mass of the perturber is $M< 2.5\times10^6$\,\Msun, making it the smallest substructure constrained by lensing above redshift 0.3. The existence of this millilens is fully consistent with the expectations from the standard cold dark matter model. On the other hand, the existence of such small substructure in a cluster environment has implications for other dark matter models. In particular, warm dark matter models with particle masses below 8.7\,keV are excluded by our observations. Similarly, axion dark matter models are consistent with the observations only if the axion mass is in the range $0.5\times10^{-22}\, {\rm eV} < m_a < 5\times10^{-22}\, {\rm eV}$.
 }
   \maketitle

\section{Introduction}
The discovery of ``MACS J1149 Lensed Star 1'' \citep{Kelly2018}, informally known as Icarus \citep{Kelly2018}, represented the birth of a new branch of astronomy dedicated to the study of luminous stars at cosmological distances (redshift $z>1$). This feat is only possible thanks to the boost provided by extreme magnification factors, $\mu$, with values  $\mu>1000$ at least for short periods of time.  In the case of Icarus, the object caught the attention of astronomers because it brightened by more than a magnitude in observations two years apart \citep{Kelly2018}. This variability was interpreted as due to a microlens---an object of approximately stellar mass in the $z=0.5444$ lensing cluster---that momentarily intersected the path of the light emitted by a $z=1.49$ background star and increased the star's observed brightness. After the discovery of Icarus, other examples of stars observed through a similar technique quickly followed \citep{Rodney2018,Chen2019,Kaurov2019}. 
The discovery of Icarus led to the prediction that JWST should find Population~III stars at $z>7$ through caustic transits \citep{Windhorst2018}. That prediction formed the basis of the JWST GTO program Prime Extragalactic Areas for Reionization and Lensing Science \citep[PEARLS:][PIDs 1176, 2738, PI Windhorst]{Windhorst2023}.

The search for lensed stars currently covers a wide range in redshift and includes objects that do not show variability. One  example is Godzilla \citep{Diego2022}, which was identified thanks to its anomalous magnification. Godzilla was identified as a monster star because of its extraordinary brightness. At $z=2.37$, it has apparent magnitude $AB \approx 22$ in the visible bands and is within reach of modest ground-based telescopes. The monster nature of Godzilla comes from the combination of three factors: i) it is extremely luminous and likely undergoing a major outburst similar to the Great Eruption of Eta Carinae in the 19th century; ii) it is close to the caustic of a powerful gravitational lens; iii) it is further magnified by a relatively large (yet invisible) substructure up to magnification factors exceeding several thousand.  The mass of this substructure is estimated to be ${\sim}10^8$\,\Msun, consistent with being a small dwarf galaxy. 
Another example is Earendel at an estimated $z=6.2$ and currently holding the record for the most distant star ever observed \citep{Welch2022a}, although Earendel may be a binary system \citep{Welch2022b}. Like Godzilla, Earendel was identified not by its change in flux but by the lack of counterimages and its proximity to the cluster critical curve. The absence of a counterimage is believed to be because the separation between the image pair is smaller than the resolving power of HST or JWST\null. The magnification of Earendel is estimated to be many thousands, making it one of the most (if not the most) extremely magnified objects known to date. Other examples of lensed stars have appeared recently in the literature \citep{Kelly2022,Diego2022,Chen2022,Meena2023a,Meena2023b}.  Several of the newly discovered stars were detected thanks to the superior sensitivity (and spatial resolution) of JWST, especially at wavelengths $>$1\,\micron, where red supergiant stars at $z>1$ are brightest \citep[e.g.,][]{Diego2023a}.

The galaxy cluster MACS J0416.1$-$403 (``M0416'')  at $z=0.397$  has been particularly fruitful for the detection of lensed stars. Two galaxies strongly magnified by this cluster contain at least four lensed stars \citep{Rodney2018,Chen2019,Kaurov2019,Kelly2022,Diego2023b}. \citep{Rodney2018} reported two 2014 transient events  consistent with being microlensed stars in a lensing arc (nicknamed the ``Spock arc''). The same arc revealed two additional 2016 transients \citep{Kelly2018}. These transients are consistent with being supergiant stars at $z\ga 1$ magnified by the combined effect of the galaxy cluster (macrolens) and stars or perhaps black holes in the intracluster light \citep[microlenses, ][]{Diego2023a}. 
More recently, \cite{Haojing2023} examined JWST/NIRCam data in four epochs separated by up to 126 days and identified 14 transients. Among these, 12 are compatible with being lensing events of distant stars. 

This paper focuses on one of these 12 lensing transients (at RA=4:16:08.84, Decl=$-$24:03:58.62) in a strongly lensed  $z=2.091$ galaxy. This peculiar object was originally identified as a lensed-star candidate in the first epoch of NIRCam observations not as a transient but as a source with extreme magnification (similar to Godzilla). The variable, or transient, nature of the source was revealed in the subsequent epochs as explained in section~\ref{sec_LS1}. The similarity between the source discussed in this paper and Godzilla lies in the fact that both are extremely luminous stars, they are both believed to be intrinsic variables, and both are highly magnified by a dark substructure that is not directly detected and with a mass in the range of dwarf galaxies. Because of this similarity, we place this new star in the category of monster (or kaiju) stars and nickname it Mothra\footnote{As explained in section~\ref{sec_LS1}, Mothra is found in the demagnification (low flux) area, similar to moths being found at night.} in order to easily refer to it throughout the paper. Alternatively,  the more official name is EMO J041608.8$-$240358, where the acronym EMO refers to Extremely Magnified Object. 

\begin{figure*} 
   \includegraphics[width=18cm]{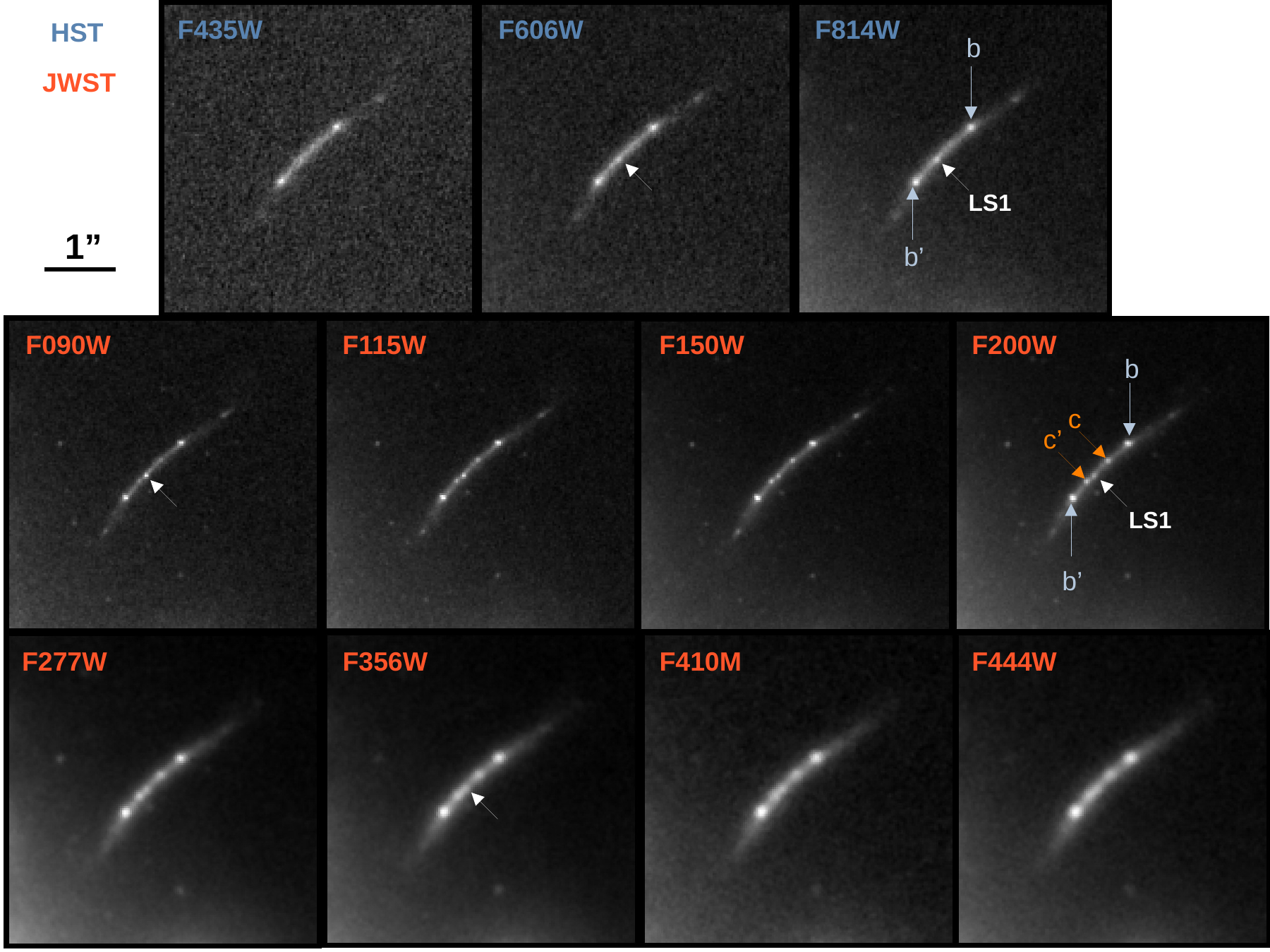}
      \caption{HST and JWST images of the arc hosting LS1.  All panels are labeled with the image's filter, and the positions of LS1 and two multiply imaged knots (b and b$'$, c and c$'$) are marked in some panels. Unlabeled arrows point to LS1. 
      A scale bar is to the left of the top row.
      The top three panels show the ACS data taken in 2014 for the HFF project.  The 2022--2023 NIRCam data (three epochs combined) are shown in the middle and bottom rows.  These are the discovery images for knots c and c$'$.
         }
         \label{Fig_Data_HST_JWST}
\end{figure*}

The paper is organized as follows.
Section~\ref{sec_data} describes the  JWST and HST data used in this work. 
Section~\ref{sec_SED} describes and interprets the spectral energy distribution (SED) of Mothra.
Section~\ref{sec_lensmodel} discusses the relevant lensing properties such as the magnification and critical curve near the lensed star. The lens model was derived using the new constraints obtained from the JWST data and is discussed in detail in a separate paper. 
Section~\ref{sec_LS1}  presents the arguments showing that Mothra must be a stellar object. 
Section~\ref{sec_globclust} shows in more depth how the alternative interpretation of Mothra being a projection effect is inconsistent with the data.
In section~\ref{sec_timevar} we discuss the observed time variability of Mothra, expected in situations where microlensing is involved or when the star is intrinsically variable. 
Further evidence in support of the lensing hypothesis is presented in section~\ref{sec_LS1prime} where we identify the possible counterimage, predicted in the lensing scenario. 
Section~\ref{sec_millilens} interprets the extreme magnification of LS1 as a possible microlensing or millilensing event. 
section~\ref{sec_maxmass}  discusses the maximum mass allowed for the millilens. 
Section~\ref{sec_discussion} discusses the results, and in particular the implications for several dark matter models; cold dark matter, warm dark matter and fuzzy dark matter. Section~\ref{sect_concl} summarizes the conclusions.  Throughout this paper, the nickname Mothra refers to the physical background source. Its observed image is referred to LS1 and its possible counterimage as LS1$'$.

In this work, we adopt a flat cosmology with $h=0.7$ and $\Omega_M=0.3$. For this model, 1\arcsec\ corresponds to 5.34\,kpc at the redshift of the cluster ($z=0.396$) and 8.44\,kpc at $z=2.091$. For the same model, the angular diameter distances that appear in the lens equation are 1102\,Mpc, 1741\,Mpc, and 1196\,Mpc for $D_d$, $D_s$, and $D_{ds}$ respectively. These are the distances to the deflector, to the source, and from deflector to the source respectively. The luminosity distance modulus for a source at redshift $z=2.091$ is 44.41. 
We use the term macrolens to refer to a lens with the approximate mass of a galaxy cluster and the term microlens for stars embedded in the intracluster light (or ICL). The term millilens is used for larger deflectors with masses comparable to dwarf galaxies. Similarly, the term caustic refers to the macromodel, and the term microcaustics refers to the microlenses. A microlens with mass 1\,\Msun\ at the redshift of the cluster deflecting light from a source at redshift  $z=2.091$ would have an Einstein radius of 2.25 microarcseconds (\mmas). Because the Einstein radius scales as the square root of the effective mass of the lens,  and the effective mass scales as the true mass times the tangential magnification, $\mu_t$,  \citep[][where the magnification  $\mu$ is the product of the tangential magnification $\mu_t$ and the radial magnification $\mu_r$]{Diego2018}, a millilens--source system with the same redshifts but with mass  $10^4$\,\Msun\ and embedded in a macromodel magnification $\mu=\mu_t\times\mu_r=250\times3.3=825$ would have an Einstein radius $\sqrt(10^4\times250)=1580$ times larger, that is $\approx 3.5$ milliarcseconds (mas). Similarly, a millilens with a mass 100 times larger (that is  $10^6$\,\Msun) would have an Einstein radius 10 times larger, that is  $\approx 35$\,mas, slightly more than the 30\,mas pixel size in JWST/NIRCam images used here. This mass range ($10^4$\,\Msun--$10^6$\,\Msun) is relevant for the discussion below.

\section{Data}\label{sec_data}

This work uses new data from the James Webb Space Telescope (JWST) together with  archival data from the Hubble Space Telescope (HST)\null. HST observed M0416 in two epochs, six months apart, in 2014 as part of the Hubble Frontier Fields program \citep[HFF, ][]{Lotz2017}. (Observations of this cluster taken in 2012 as part of the CLASH program \citealt{Postman2012} are too shallow for our purposes and not used here.)  Data used here are from the ACS filters F435W, F606W, and F814W and reach depths from 28.8\,mag in F435W to 29.1\,mag in F814W\null. JWST/NIRCam observed the cluster in four epochs in 2022--2023 over a time span of about four months. Epochs 1, 2, and 4 are from the PEARLS program \citep{Windhorst2023}, and Epoch 3 is from the CANUCS program \citep{Willott2022}.  Each epoch included observations in eight filters: F090W, F115W, F150W, F200W, F277W, F356W, F410M, and F444W\null. 
The PEARLS epochs reached depths ($5\sigma$ point source limits in AB mag) of 28.7, 28.8, 28.9, 29.1, 28.7, 28.8, 28.2, and 28.5, respectively, while the CANUCS epoch is about 0.2--0.3\,mag deeper.
The three PEARLS epochs combined  reach $\sim$0.6\,mag deeper than the individual epochs.
\citet[][their Table~1]{Haojing2023} provide exact dates,  exposure times, and depths reached in each epoch and each filter. This work used images with a pixel scale of 30\,mas.

The strongly lensed arc that is the focus of this paper (RA=64.03675, DEC=$-$24.06624) was identified by the HFF program as a lensed galaxy. This arc has two discrepant spectroscopic redshift estimates obtained from MUSE data, $z=1.827$ \citep{Richard2021} and $z=2.091$ \citep{Bergamini2021}. We adopted $z=2.091$ because it is closer to the redshift predicted by the lens model \citep{Diego2023c}, and it is consistent with the position of the previously unidentified third counterimage, found now in the new NIRCam data. The higher redshift is also more consistent with the photometric redshift estimate $z=2.23$\footnote{http://cosmos.phy.tufts.edu/~danilo/HFF/Home.html}.

Figure~\ref{Fig_Data_HST_JWST} shows HST and JWST images of the arc. Each multiply lensed feature is labeled with a single letter, and the counterimage of that feature has a prime. Mothra is in the middle of the arc. The 2014 HST data show Mothra as well as knots b and b$'$. Knots c and c$'$ are seen only in JWST data and are key for the interpretation because they constrain the position of the critical curve to be close to the midpoint between c and c$'$ independent of any specific lens model. 
The new knots c and c$'$ emphasize the anomaly of LS1 because a counterimage, LS1$'$, is expected between LS1 and c but is not seen yet  LS1 was visible in 2014.  If LS1$'$ is missing because the source is transient and the two paths have different light travel times, the time difference must be $>$8~years.

\begin{figure} 
   \includegraphics[width=8.5cm]{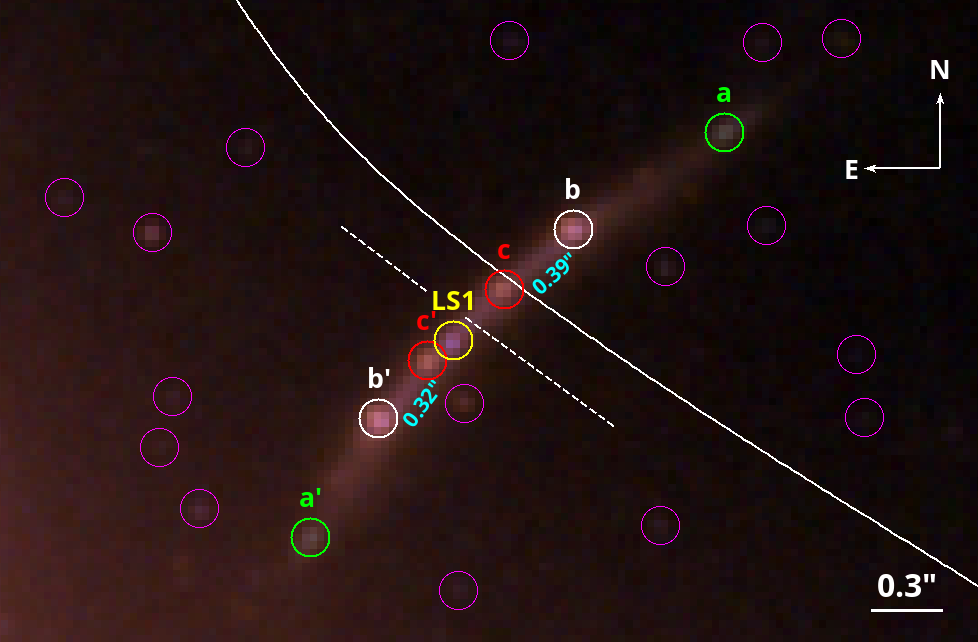}
      \caption{
    Enlarged color image of Mothra's arc and its surroundings. Colors are a combination of HST and JWST filters; F435W+F606W+F814W+F090W+F115W, F150W+F200W, and F277W+F356W+F410M+F444W for blue, green and red respectively.
    LS1 and three multiply lensed knots and their counterimages are labeled. No counterimage for LS1 is visible. Numerous faint, unresolved objects are marked with unlabeled magenta circles. These could be globular clusters or compact galactic remnants in the galaxy cluster. The white dashed line is the inferred position of the critical curve based on the ratio of the b--c separation (0\farcs39) to the b$'$--c$'$ separation (0\farcs32). The solid white curve is the expected position of the critical curve based on the lens model. The two curves are ${\approx} 0\farcs3$ apart. 
         }
         \label{Fig_Star}
\end{figure}

\section{SED fitting}\label{sec_SED}
Photometry of LS1 is complicated by its location in a strongly lensed arc and by the nearby point source c$'$ as shown in Figure~\ref{Fig_Star}. At longer wavelengths, the instrument PSF blends LS1 and c$'$ (Fig~\ref{Fig_Data_HST_JWST}). To overcome these limitations, we fit a point-spread function (PSF) to LS1 and to each of the six nearby knots in the three-epoch combined image. The PSF model was derived from the stacked signal of nearby, unsaturated, unresolved sources on the same image.  For the filters with $\lambda>1.5$\,\micron, in which LS1 and c$'$ are partially blended, we subtracted a model of the arc prior to PSF fitting. The model was scaled from the residual in the F150W band after point source subtraction and smoothed to match the resolution of the longer wavelength filters. Details are given in the Appendix.

\begin{figure} 
   \includegraphics[width=9cm]{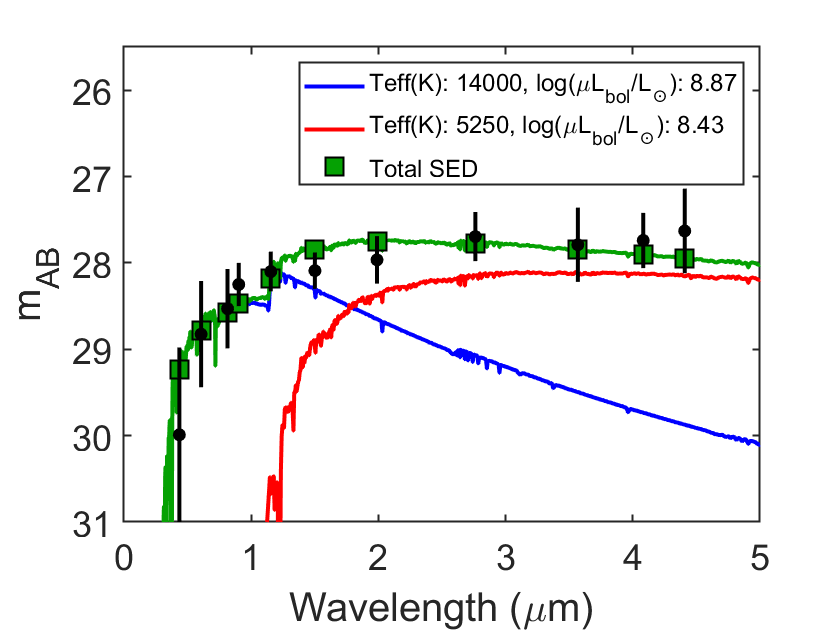} 
      \caption{
Mothra's SED and the best matching binary model. The blue and red lines represent model stellar spectra for stars with $T_\mathrm{eff}$ and $\mu L_\mathrm{bol}$ as shown in the legend.  Models are from the \citet{Lejeune97} compilation of stellar atmosphere spectra at solar metallicity redshifted to $z=2.091$. The green line shows the combined SED\null. Black circles with error bars represent the observed photometric data and green boxes the model photometry resulting from the best-fit compound spectrum. If the two stars experience the same magnification, the high--$T_\mathrm{eff}$ star has a bolometric luminosity $\approx$0.4\,dex (a factor of $\approx 2.5$) higher than that of the low--$T_\mathrm{eff}$ star.}
         \label{Fig_SED_LS1}
\end{figure}

The measured SED for LS1 is shown in Figure~\ref{Fig_SED_LS1}. The SED is too broad to come from a single star, but a binary system with temperatures $T\approx 14000$\,K and $T\approx 5250$\,K matches well.  Finding a binary is not surprising because most massive stars in our Galaxy are binaries, and the binary
fraction of massive stars seems to go up at lower metallicities \citep[see e.g., the discussion of][for some references on this]{Windhorst2018}.
%
The upper limit of likely magnifications requires the stars to be massive (initial masses $\gtrsim 10$\,\Msun), and therefore only low surface gravities need to be considered. Dust reddening would demand higher luminosities, and we assume zero. The best fit is a hot star with $T_\mathrm{eff}=14000$\,K and $\log(\mu L/L_\odot)\approx 8.9$ plus a cool star with $T_\mathrm{eff}=5250$\,K and $\log(\mu L/L_\odot)\approx 8.4$. For an adopted magnification of $\mu\approx 5000$, the intrinsic luminosity of the cooler star would be $\log(\mu L/L_\odot)\approx 4.7$ , which would correspond to a yellow super/hypergiant star inside the instability strip, with initial mass $M\approx 15\ M_\odot$ \citep{Ekstrom12,Szecsi22}. If the two stars experience a similar magnification, then the higher-$T_\mathrm{eff}$ B-type star must have an intrinsic luminosity a factor of $\approx 2.5$ higher. A red supergiant ($T_\mathrm{eff} \approx 4000$\,K) would provide a better SED fit, but Mothra exhibits significant flux variations at 1.5\,\micron\ and longer wavelengths, yet not at 0.9 or 1.15\,\micron\ (Figure~\ref{Fig_TimeVar_All}). This means the red component is varying, and it has to contribute significant light to the F150W band.  This requires $T_\mathrm{eff}\ga5000$\,K.
At peak brightness, Mothra becomes redder \citep{Haojing2023}, which requires the cooler component to transition into the red supergiant regime or alternatively experience a boost in luminosity accompanied by increased circumstellar dust reddening. 

Under the assumption of single-star evolution and similar magnifications for the two components, their luminosity ratio creates an apparent age discrepancy, since the more luminous star is expected to evolve into the $T_\mathrm{eff}<20000$ K regime significantly ahead of the lower-$L$ star, whereas both stars here seem to be observed in these short-lived states. Due to the degeneracies involved in this two-component fit, the current SED does not allow dust effects towards the two stars to be meaningfully constrained, but the effect of significant dust attenuation and reddening towards either star would raise it's $T_\mathrm{eff}$ and intrinsic bolometric luminosity beyond what is inferred from the current fit. In a scenario where the low-$T_\mathrm{eff}$ star is more strongly affected by dust, as could happen in the case of significant circumstellar dust around red or yellow supergiants (e.g., \citealt{Massey05,Drout12}, but see \citet{Beasor22} for a different view), the luminosity gap between the two stars could be reduced or even reverted.

Another possibility is that the bluer, higher-$L$ star is the rejuvenated result of a stellar merger, with extended lifetime as a result \citep{Glebbeek13,Schneider16}. Merger products of this type have recently been proposed to explain the extended main-sequence turnoff problem of young star clusters \citep{Wang22}, in which massive stars with seemingly discrepant ages seem to co-exist in the same cluster. While $L_\mathrm{high\ T_{eff}}>L_\mathrm{low\ T_{eff}}$ pairs of stars do not appear in local samples of red supergiant binaries \citep{Patrick22}, our constraints on the source size of Mothra do not necessarily require the two stars contributing to the SED to be a binary pair, as they could just be two bright members of the same star cluster.

Yet another potential solution is that the shorter-wavelength, non-varying part of the SED is dominated not by a single star, but by the integrated light from many young stars in the star cluster that hosts the varying yellow/red supergiant. Due to its much larger source size (parsec-scale), the star cluster would be more moderately magnified ($\mu$ in the hundreds) than the yellow/red supergiant ($\mu$ in the thousands), and would still allow Mothra to appear point-like. This solution would not cause any apparent age discrepancy between the two components, since the most luminous blue star in the cluster would then have a bolometric luminosity similar to or below that of the red supergiant. A young, moderately-magnified star cluster (age $\lesssim 20$ Myr, mass $\lesssim 10^5 \ M_\odot$) plus a highly magnified supergiant star could also explain the overall SED shape of Mothra, but only under the assumption of significant dust attenuation and reddening ($A_V> 0.5$ mag), since the young star cluster would otherwise not be able to reproduce the red slope of the photometric SED at wavelengths $\lesssim 1 \mu$m. The main problem with scenarios of this kind is that it is difficult to find a solution where this young star cluster would dominate the light at $\leq 1.15\ \mu$m, where no significant variabilitiy is seen, yet at the same time not completely outshine the star at $1.5\ \mu$ m, where Mothra is seen to vary substantially in brightness (Figure~\ref{Fig_TimeVar_All}). As we have been unable to find a satisfactory solution of this kind, we consider this scenario less likely than the two-star explanation for the properties of Mothra.

\section{Lens Model}\label{sec_lensmodel}
\citet{Diego2023c} describes a new lens  model for M0416. The model uses previous lensing constraints from HST \citep{Zitrin2013b,Jauzac2014,Diego2015,Caminha2017,Richard2021,Bergamini2021,Diego2023b} and a new set of constraints derived from the PEARLS JWST data.
Only the small area containing the arc around Mothra (Figure~\ref{Fig_Data_HST_JWST}) is relevant for the present paper. The reader can find the full details of the new lens model in \citet{Diego2023c}.

Figure~\ref{Fig_Star} shows the strongly lensed  $z=2.091$ galaxy that hosts Mothra.
The galaxy is imaged twice, forming an elongated arc with the cluster-lens critical curve (white solid line) passing close to the midpoint of the arc.  With lensing constraints from b--b$'$ but not c--c$'$, the model critical curve is offset from the c--c$'$ midpoint by $\approx$0\farcs3. Such offsets between predicted and observed positions are typical in lens models.  For this location in particular, the proximity of the northern BCG   makes the lens model less accurate because the BCG contributes significantly to the lensing potential, but this BCG has no radial arcs near it to determine an accurate mass for it.

\begin{figure} 
    \includegraphics[width=8.5cm]{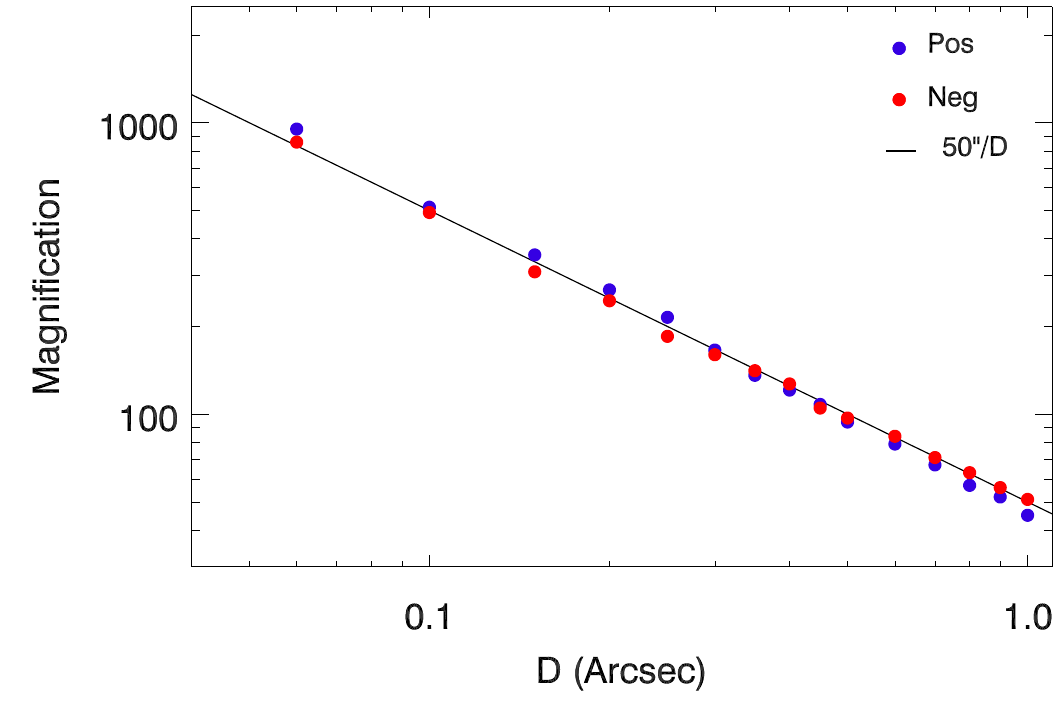}
      \caption{Magnification versus distance. The solid black line represents the law $\mu=A/D$, with $D$ being the distance to the critical curve expressed in arcseconds and $A=62"$. The blue and red points are measured magnification values from our lens model, at the position of the critical curve intersecting the arc and in a direction perpendicular to the critical curve. For the critical curve we assume the is at $z=2.091$. The blue points correspond to magnification values measured on the side with positive parity (north-west of the white solid curve in Figure~\ref{Fig_Star}) and the red points are for magnifications measured on the side with negative parity (south-east of the same curve). At a distance of $D=0.07"$, the black curve predicts magnification $\approx 8855$. 
         }
         \label{Fig_Mu_vs_D}
\end{figure}

A more precise estimate for the position of the critical curve can be obtained from basic lensing principles. 
Galaxies that intersect cluster caustics form arcs with distinctive features that repeat on either side of the critical curve. The magnification decreases as one moves away from the caustic, usually following a well defined law $\mu=A/D$ where $D$ is the distance to the critical curve expressed in arcseconds, and $A$ is a normalization constant that depends on the lens model. Different portions of the critical curve have different values of $A$. The largest values of $A$ are often found on elliptical lenses at the extremes of their critical curves. These extremes correspond to the cusps of the caustics in the source plane.
Near the critical curve, the law $\mu=A/D$ is relatively accurate, only departing from it when $D\ga 1\arcsec$. For smaller separations, the critical curve is near the midpoint between an image pair.  A more precise estimate can be obtained when there are two image pairs by taking magnification into account. Here the ratio of separations $\hbox{b--c}/\hbox{b$'$--c$'$} = 0.39/0.32=1.22$ gives a good approximation for the magnification ratio on the two sides of the critical curve. Applying this ratio to the c--c$'$ separation (0\farcs43) puts the critical curve $0\farcs19$ from c$'$ (and 0\farcs24 from c) as marked in Figure~\ref{Fig_Star}.  This is 0\farcs07 from LS1. Images in the region to the north-west of the critical curve have positive parity (that is, similar orientation to the unlensed image) while images appearing to the south-east of the critical curve have negative parity. In the absence of micro- or millilenses, the magnification of both negative and positive parity images is often very similar as shown in Figure~\ref{Fig_Mu_vs_D}. When micro- or millilenses are present, the magnification of microimages corresponding to very small sources (such as stars) can be very different depending on the parity.  Microimages of lensed stars with positive parity are often magnified (with respect to the macromodel magnification) by micro- and millilenses while microimages with negative parity are often demagnified (with respect to the macromodel magnifcation) by small deflectors (micro- and millilenses). An important takeaway from this section is the fact that LS1 is observed on the side with negative parity, so demagnification is expected, yet LS1 is observed but not its counterimage LS1$'$ which has positive parity and hence expected to often have more magnification that the one predicted by the macromoel. This issue will come later on when we discuss the interpretation of Mothra. A major difference between LS1 and other extremely magnified stars such as Earendel is that LS1 is not on the critical curve.  The implication is that there should be a counterimage of LS1 $\approx$0\farcs7 on the other side of the critical curve, but no such image is obvious in the current data. \\

Another important parameter is the magnification at the position of LS1. 
The \citet{Diego2023c} model gives normalization factor $A=62\arcsec$ at the position of the $z=2.091$ arc. With LS1 0\farcs07 from the critical curve, $\mu \approx 885$. LS1 appears unresolved, and the magnification sets an upper limit on source size $R<1$\,parsec. A compact group of stars such as R136a would fit the size and luminosity constraints but cannot explain the lack of counterimage. To explain this, the source must be very luminous and magnified by extreme values on one side of the critical curve, while on the other side its counterimage is magnified by a factor close to the most likely value from the macromodel, $\mu \approx 885$. This situation is similar to Godzilla, where among the several counterimages predicted, only one is observed \citep{Diego2022} thanks to the extra magnification provided by a nearby millilens. The lack of counterimage detection at 29.5\,mag together with $\mu \approx 885$ gives an upper limit on the source's intrinsic flux density corresponding to $\approx$36.9\,mag and absolute magnitude fainter than $-7.5$. Supergiant stars fall into this category and are a prime candidate to explain Mothra.

The detection of LS1 at $\approx$28\,mag requires a boost in magnification of at least $\approx$1.5\,mag above the magnification of the counterimage. Furthermore, this boost must have been maintained over at least 8 years because the source was already visible in 2014. If microlensing is involved to explain LS1,  high sustained magnification can be obtained only if the star is moving nearly parallel to the direction of the microcaustic and very close to it. That is, such a scenario would require a very high degree of fine tuning between the relative direction of motion and the orientation of the microcaustic. This scenario is explored in more detail in section~\ref{sec_millilens}. 


Finally, another important variable from our lens model is the value of the radial magnification, which to first order can be considered more or less constant along the arc. The lens model predicts for the convergence and shear, $\kappa\approx 0.85$ and $\gamma \approx 0.15$ respectively, at the intersection between the critical curve and arc. This results in a radial magnification factor of $\mu_r\approx 3.3$. The tangential magnification changes rapidly as one moves away form the critical curve following the canonical $\mu \propto D^{-1}$.

\section{An extremely magnified and compact source}\label{sec_LS1}

Although we have already established the possible interpretation of LS1 as a binary star undergoing extreme magnification, before proceeding any further it is imperative we consider other more mundane possible interprations. LS1 and the absence of a counterimage can be interpreted in only a few ways:
\begin{enumerate}[label=\roman*)]
\item LS1 is a transient event, and due to time delays, we have not yet seen the transient's counterimage, or alternatively, the counterimage has already disappeared.
\item There is in fact an image pair, but the two are so close to each other that they appear as an unresolved image. This scenario would be similar to  Godzilla or Earendel. 
\item The source is a foreground object, and we are seeing a projection effect.
\item  Microlensing is temporarily increasing the flux of LS1 but not its counterimage.
\item  Millilensing is temporarily increasing the flux of LS1 but not its counterimage. The difference from microlensing is that the timescale would be longer.
\end{enumerate}
The next subsections explore these possibilities.

\subsection{LS1: A transient event?}
Among the possible scenarios above, i) can be easily discarded because the lens model gives a time delay between b and b$'$ between 120 days (for a lens model that uses only the available spectroscopically confirmed systems) and 230 days (for a lens model that in addition uses the newly discovered JWST lensed systems, and with estimated geometric redshifts from the lens model). For LS1 and its counterimage the time delay, would be even shorter because the separation between them is smaller than the separation between b and b$'$. Because LS1 has been observed for at least 8 years, any intrinsic change in flux that makes it detectable on one side of the critical curve should make it detectable on the other side as well within the time frame covered by our JWST observations.  The same argument applies even for the NIRCam observations alone, which span 126 days. 

\subsection{LS1: An unresolved pair of images?}
A key element to interpret the different possibilities is the precise position of the critical curve in relation to the source. As discussed in tye previous section, the critical curve can not be at the position of LS1 but it must be offset by $\approx 0\farcs07$. Its counterimage should be found at a similar distance form the critical curve but in the opposite direction.  Finding LS1  0\farcs07  from the critical curve (Section~\ref{sec_lensmodel}) rules out scenario ii). 
The fact that the source is found on the side with negative parity is not surprising {\it per se}. Microcaustics on the  negative-parity side are in general more powerful than the corresponding caustics on the positive-parity side (for the same macromodel magnification and microlens mass). This is consistent with flux conservation arguments, since the more powerful caustics for images with negative parity compensate the areas of demagnification which are present on this side of the critical curve, but do not exist on the side with positive parity. What is more surprising is that the magnification must have remained nearly constant for at least 8 years. Comparison between the 2014 F160W ACS data and the 2022--2023 F150W NIRCam data shows no evidence of flux variation at the position of LS1. However, this test is limited by the relatively low resolution of HST compared to JWST\null. Section~\ref{sec_timevar} addresses the variability of LS1 during the four NIRCam epochs. 


\subsection{LS1: A globular cluster or dwarf galaxy?}\label{sec_globclust}

\begin{figure} 
   \includegraphics[width=8.5cm]{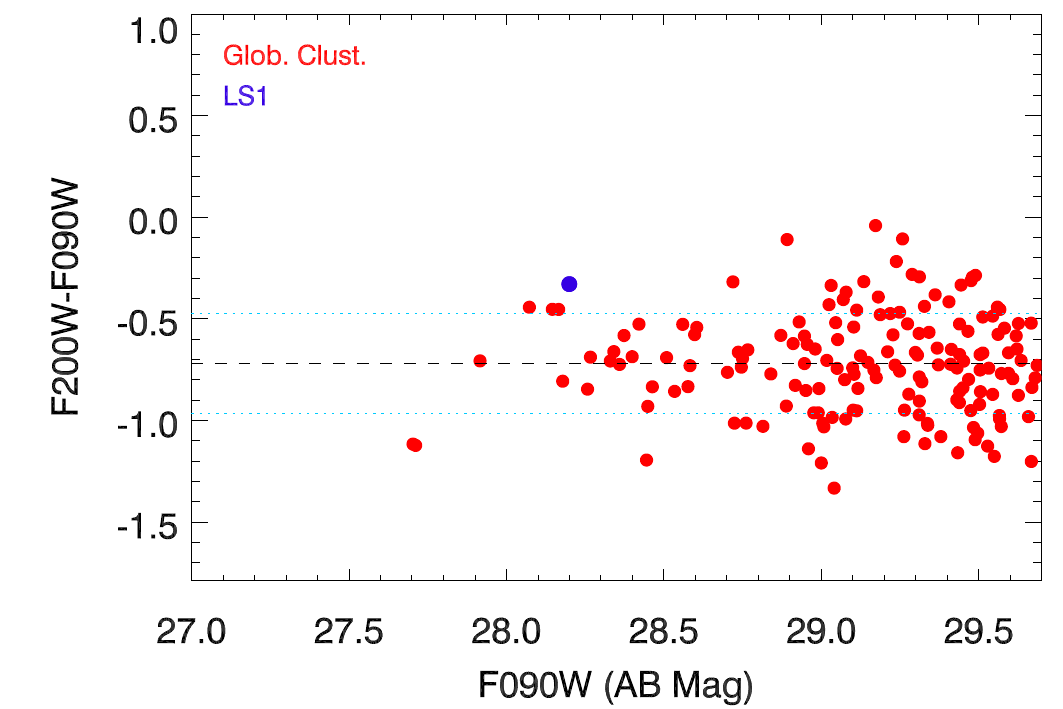}   
      \caption{Color-color plot for the globular clusters. The red points correspond to globular clusters found in the central regions of M0416, and near LS1. The blue dot is for LS1. 
      Magnitudes are computed after fitting the star-based PSF model. LS1 is clearly an outlier when compared with the globular clusters.  
      Errors in the magnitudes have been omitted for clarity purposes. 
      The horizontal dashed is the mean F200W-F090W of the globular clusters. The blue dotted ones represent 1 standard deviation. 
         }
         \label{Fig_GlobularClusters}
\end{figure}

As shown in Figure~\ref{Fig_Star}, the LS1 arc  is surrounded by several compact sources. These could be globular clusters or remnant galactic cores that have had their outer envelopes tidally stripped, but for simplicity we will refer to them as GCs. In the area shown in the figure, the number density of GCs is $\approx$1.6\,arcsec$^{-2}$. The area between c and c$'$ (the separation between c and c$'$ times the arc's thickness) is $\approx$0.05\,arcsec$^{-2}$. The probability of a compact source falling in this region is thus $\approx$8\%. 
Although relatively small, this probability is  large enough that the possibility of LS1 being a GC in M0416 needs to be considered seriously. 

LS1 is brighter than any of the GCs, making the probability that it is one considerably smaller than 8\%. LS1 is also bluer than the nearby GCs. This is evident in Figure~\ref{Fig_Data_HST_JWST}, where none of the GCs is detected in HST's F606W, and only one can be  seen in F814W\null. 
Figure~\ref{Fig_GlobularClusters} gives more quantitative comparison via a color--magnitude plot. LS1 stands out as the bluest object with $\rm F090W < 28.5$\,mag.
Although this suggests LS1 is not a GC, it does not entirely rule out its being an object along the line of sight, for instance in M0416 (and hence not magnified).  


\begin{figure} 
   \includegraphics[width=\linewidth]{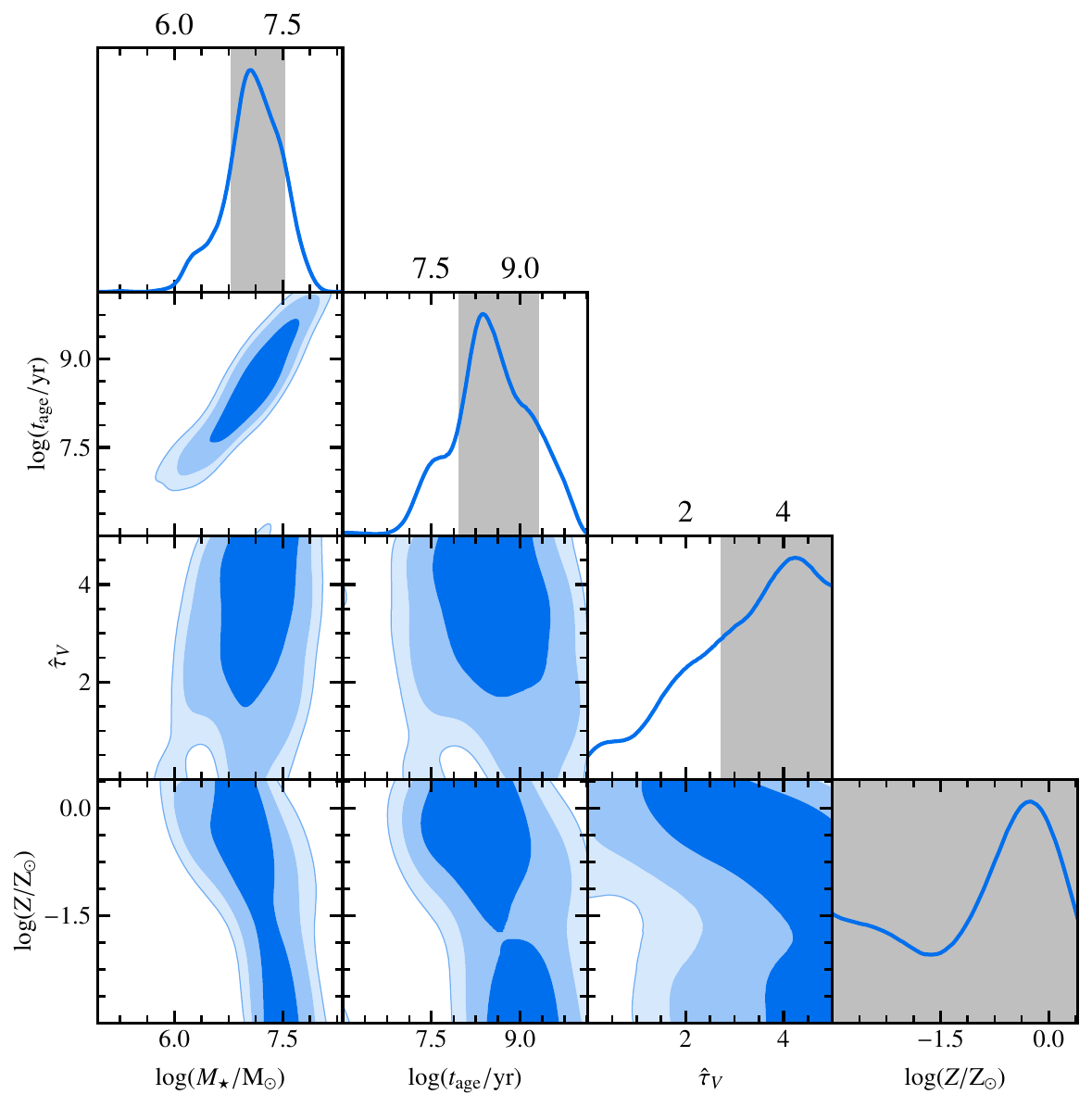}   
      \caption{Posterior distributions of the four \texttt{BEAGLE} SSP fit parameters assuming Mothra is a globular cluster at $z=0.396$. The four derived parameters are stellar mass $M_{\star}$, stellar age $t_{\mathrm{age}}$, dust attenuation optical depth ${\tau}_V$, and stellar metallicity $Z$. The blue contours represent the 68\,\%, 95\,\% and 99\,\% contours of the full joint posterior distribution respectively and the grey shaded are represents the (marginalized) 68\,\% interval of each individual parameter.}
         \label{Fig_GC_Beagle}
\end{figure}

If LS1 is one of the many globular clusters observed near it, the SED will reveal the stellar population parameters.  We fit the observed SED with the \texttt{BayEsian Analysis of GaLaxy sEds} tool \citep[\texttt{BEAGLE}:][]{Chevallard2016}. The fit used the \citet{Bruzual2003} stellar population models with a \citet{Chabrier2003} initial mass function (IMF) and a single stellar population (SSP) at fixed redshift $z=0.396$. Figure~\ref{Fig_GC_Beagle} shows the posterior distributions of the fit parameters. If LS1 is indeed a globular cluster, it must be massive ($\log(M_{\star}/\mathrm{M}_{\odot})>6.8$) and  heavily dust-obscured ($\tau_V > 2.7$).  These properties are more consistent with a tidally stripped galaxy nucleus than a globular cluster.
That would make the mass even higher because a galaxy nucleus, as a compact stellar structure with large escape velocities, should contain a non-negligible amount of dark matter. A fundamental problem with this picture is that a mass as high as ${\approx}10^7$\,\Msun\ would change the relative magnifications of image pairs c and c$'$. This is not observed: these two knots have flux ratios close to unity, ruling out the hypothesis that LS1 is a ${\approx}10^7$\,\Msun\ stellar structure at $z=0.396$. 

LS1 could have a lower stellar mass if it is less distant than M0416. A globular cluster at $z=0.12$, where a few galaxies are found in the  field, would have stellar mass ${\approx}10^6$\,\Msun\ and would not unduly distort the c/c$'$ magnification ratio.  The problem is that  the fit  of this model to the LS1 SED is poor. In particular, the observed flux densities at $\lambda >3$\,\micron\ are $\approx$0.85\,mag above the observed ones. (At low redshift, rest-frame and observed wavelengths are nearly the same, and globular clusters do not exhibit infrared excesses.)
On the opposite direction, if LS1 is a foreground object behind the cluster but in front of the lensed galaxy, its mass needs to be even larger in order to fit the photometry, worsening the lensing constraints from c and c'. Only if the redshift of LS1 is very close to the lensed galaxy (but below it), could LS1 appear as not multiply lensed and have a small lensing effect on the background galaxy. But this would require an extraordinary (and possibly entirely new) type of object since it would also be extremely magnified yet unresolved. It must be also very dust-obscured in order to explain the observed SED. Such object should not show time variability or have a counterimage, two features typical of lensed stars which we discuss in more detail in sections~\ref{sec_timevar} and~\ref{sec_LS1prime}\footnote{An AGN type-of-object could show variability but more in the blue tan in the red component, opposite to what we find in section~\ref{sec_timevar}. Also, reddening would be incredibly high given the typical blue nature of this type of objects.} 
All in all, the possibility of LS1 being a foreground object is ruled out.


\subsection{LS1 as a microlensing event}\label{sec_microlens}

Microlenses are expected to be present in relatively large numbers at LS1's position near  the northern BCG in M0416. However, stellar microlenses produce relatively small caustics that can boost the magnification by the necessary factor ($\ge$4 with respect to the macromodel magnification) only for small periods of time (usually weeks to months for typical relative velocities). 
A possible solution to this small time scale is if the background star is moving paralell to a microcaustic. This would boost the image with negative parity (LS1) during more extended periods while leaving the counterimage with positive party (LS1$'$) undetected. 

\begin{figure} 
      \includegraphics[width=8.5cm]{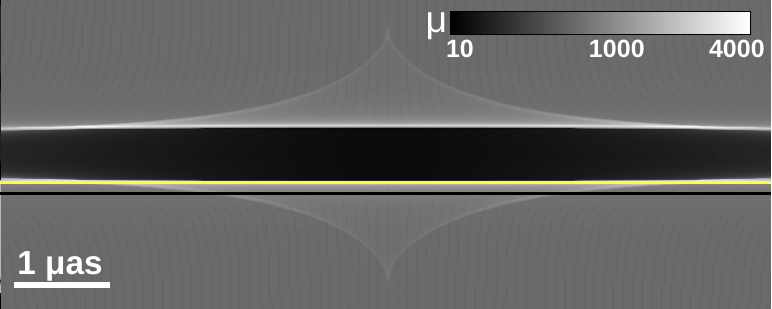}
      \caption{Caustics around a 1\,\Msun\ microlens on the side with negative parity at the redshift of the cluster lensing a  background source at $z=2.091$. The macromodel magnification in this region is $\approx$850, and the gray scale indicates the combined magnification $\mu$ of the macromodel and the microlens. The black horizontal band corresponds to the demagnification region that exists only for images with negative parity. The microlens can demagnify a lensed star down to  $\mu\approx 10$ in this band. Lighter triangular regions above and below show high-magnification regions.  The yellow horizontal line illustrates a trajectory grazing the bottom horizontal caustic, and the black line is for a trajectory 0.1\,\mmas\ south of the yellow line. These lines correspond to the yellow and black curves in Figure~\ref{Fig_Microlens2}. 
         }
         \label{Fig_Microlens1}
\end{figure}

In order to test the microcaustic hypothesis, we simulated a microlens near the critical curve on the side with negative parity. We adopted values for the tangential and radial magnification  $\mu_t=250$ and $\mu_r=3.3$, consistent with the lens model. This results in a total magnification $\mu=825$. For the microlens, we adopted a mass of 1\,\Msun. Most microlenses in the intracluster light are expected to be lighter, but a few could be even more massive. More massive objects include also remnants (neutron stars and black holes) and could potentially include massive but compact candidates for dark matter such as primordial black holes. A mass of 1\,\Msun\ offers a good compromise between all these scenarios. Figure~\ref{Fig_Microlens1} shows the magnification and caustics from this microlens from a standard ray-tracing technique at spatial resolution of 10 nanoarcseconds per pixel.  The macrocaustic at LS1's position is smooth: a nearly horizontal line a few parsecs (equivalent to a few hundred microarcscec) away from the microlens.

\begin{figure} 
   \includegraphics[width=8.5cm]{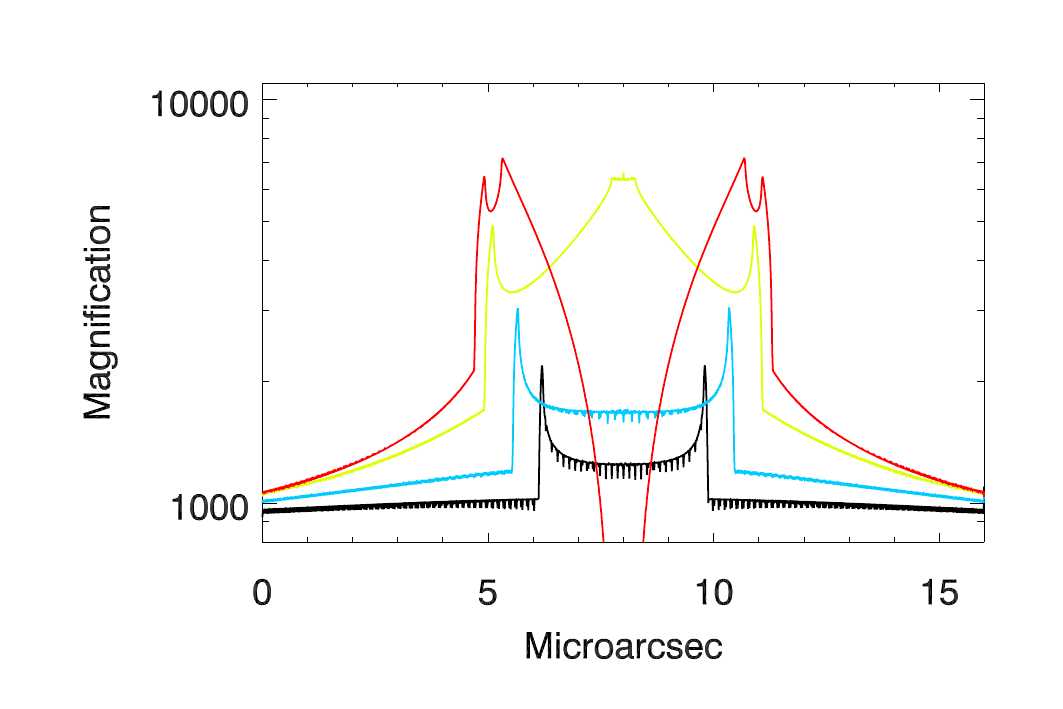}
      \caption{One dimensional magnification (light curve) of a star moving in a horizontal direction across the microlens shown in Figure~\ref{Fig_Microlens1}.  The black and yellow curves correspond to the tracks shown in yellow and black respectively in  Figure~\ref{Fig_Microlens1}. The track for the blue curve is also horizontal and is midway between the black and yellow tracks. The red curve corresponds to a track one pixel (10\,nanoarcsec) above the yellow track, where the trajectory intersects both the caustic and the area of demagnification. The radial magnification from the macromodel is 3.3, and the tangential one is 250. At large distances from the microlens, the magnification converges to the macromodel value  $\mu=825$. The small scale fluctuations are pixel noise from the simulation.
         }
         \label{Fig_Microlens2}
\end{figure}

The modeled caustics form two triangular shapes (Figure~\ref{Fig_Microlens1}).  The largest magnification factors are found near the cusps of the caustics in two regions $\approx$10\,\mmas\ in maximum size.  Because the microlens is on the side with negative parity, there is a central $\approx$1\,\mmas\ region adjacent to the caustics that can demagnify sources by several magnitudes. In order to  be at least one magnitude brighter than the counterimage, the source must be close to the region of maximum magnification. Figure~\ref{Fig_Microlens1} illustrates two trajectories for a source moving parallel to the demagnification zone, and Figure~\ref{Fig_Microlens2} shows the magnifications, i.e., light curves, corresponding to these trajectories and to two others.
While it is possible for a 1\,\Msun\ microlens to maintain a nearly constant, high magnification for eight years ($\approx$1\,\mmas\ if the relative velocity $v_r=1000$\,km\,s$^{-1}$), to do that, the microlens' relative motion must be perfectly aligned with one of the microcaustics.
The degree of fine tuning required makes this scenario unlikely. 

The requirements on the direction of motion of the microlens can be relaxed if the relative velocity is smaller or if the microlens is more massive. A smaller relative velocity keeps the source in the high-magnification region longer. For the redshifts of the lens and background source, one could consider smaller values for the velocity by a factor $\approx$2, but this requires fine tuning between the relative motions of the observer, lens, and source. Due to the high degree of fine tuning required to explain LS1 as a microlensing event we deem this possibility as unlikely. We conclude this section by leaving the microlensing hypothesis as the only viable possibility to explain LS1 as a long duration anomaly in the flux of a lensed star.

Increasing the mass of the perturber allows a wider range of distances between the source and microcaustic because the strength of the microcaustic scales with the mass of lens. This also eliminates the need for fine tuning in the relative direction of motion since multiple trajectories near the caustic of the perturber can maintain the needed magnification for at least 8 years.  The next two sections present additional evidence in favor of millilensing, and Section~\ref{sec_millilens} describes the millilens scenario in  detail. 

\section{Time variability}\label{sec_timevar}
If LS1 is a small galaxy or globular cluster along the line of sight, it would not vary in flux unless it hosts a variable source such as an AGN\null. If, on the contrary, LS1 is a small source such as a (binary) star, variability is expected because supergiant stars (especially cool ones) are often variable \citep{Kiss2006,Yang2011}. Also, microlensing from stars in the intracluster medium (ICM) should cause small changes in observed flux over time. At macromodel magnification factors of order 1000, as expected for LS1, even a modest mass density of microlenses of a few \Msun\,pc$^{-2}$ should produce observable variations \citep{Diego2018}. 
Also, as shown by \cite{Diego2023b}, M0416 is expected to lens a wealth of $z>1$ red supergiant stars that happen to fall near cluster critical curves. These stars are often intrinsic variables. 

\begin{figure} 
  \includegraphics[width=8.5cm]{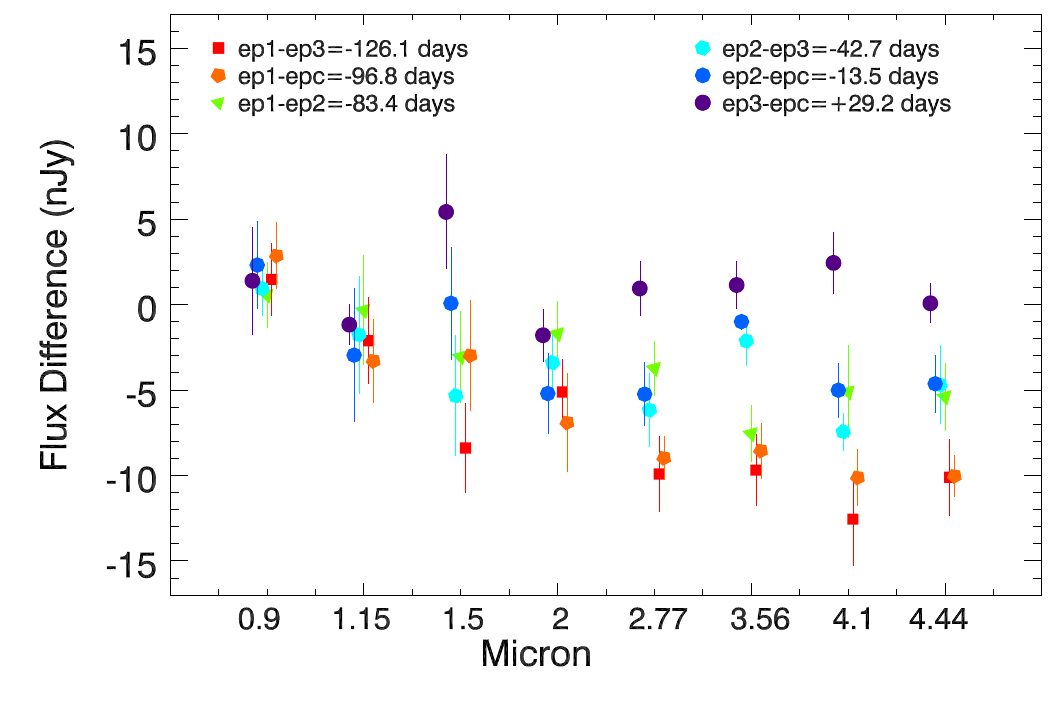}
      \caption{Time variability of LS1. The y-axis shows the difference in flux between different epochs in apertures of radius 0\farcs09 centered on LS1 (error bars are 1-$\sigma$ derived from random positions outside the arc). Each color corresponds to a different combination of epochs, as indicated by the legend inside the figure.  The legend gives the time of each epoch in days after Ep1.
         }
         \label{Fig_TimeVar_All}
\end{figure}

To search for time variability of LS1, we measured its flux in fixed apertures of 0\farcs09 radius on each of the four single-epoch images and also on difference images described in Appendix~\ref{sec_CornerPlot}. PSF subtraction was not used here because aperture photometry is less sensitive to position-angle differences between epochs, and the non-variable arc contamination does not matter. \citet[][their Table~4]{Haojing2023} did photometry by a different method, and the results are consistent. Figure~\ref{Fig_TimeVar_All} shows LS1's flux changes in the four epochs. A clear trend with time is apparent with LS1 showing a large increase in brightness from Ep1 to Ep2 and then smaller increases to Epc and then Ep3.  The changes are largest in the F410M filter, where the change from first to last epoch is 5$\sigma$.  Similar but smaller fluctuations are observed at  1.5\,\micron\ and longer wavelengths. In contrast, there are  no significant variations at 0.9 or 1.15\,\micron.
This chromatic effect can be easily understood if the source is a binary star (or an unrelated pair of stars at separations of 1\,parsec or less) with a blue and red component, and the red component is intrinsically varying in flux. The variation is about 0.65\,mag \citep{Haojing2023} in F410W and F444W (rest wavelengths 1.3--1.4\,\micron) in 41 rest-frame days.
 
%

\begin{figure} 
    \includegraphics[width=9.0cm]{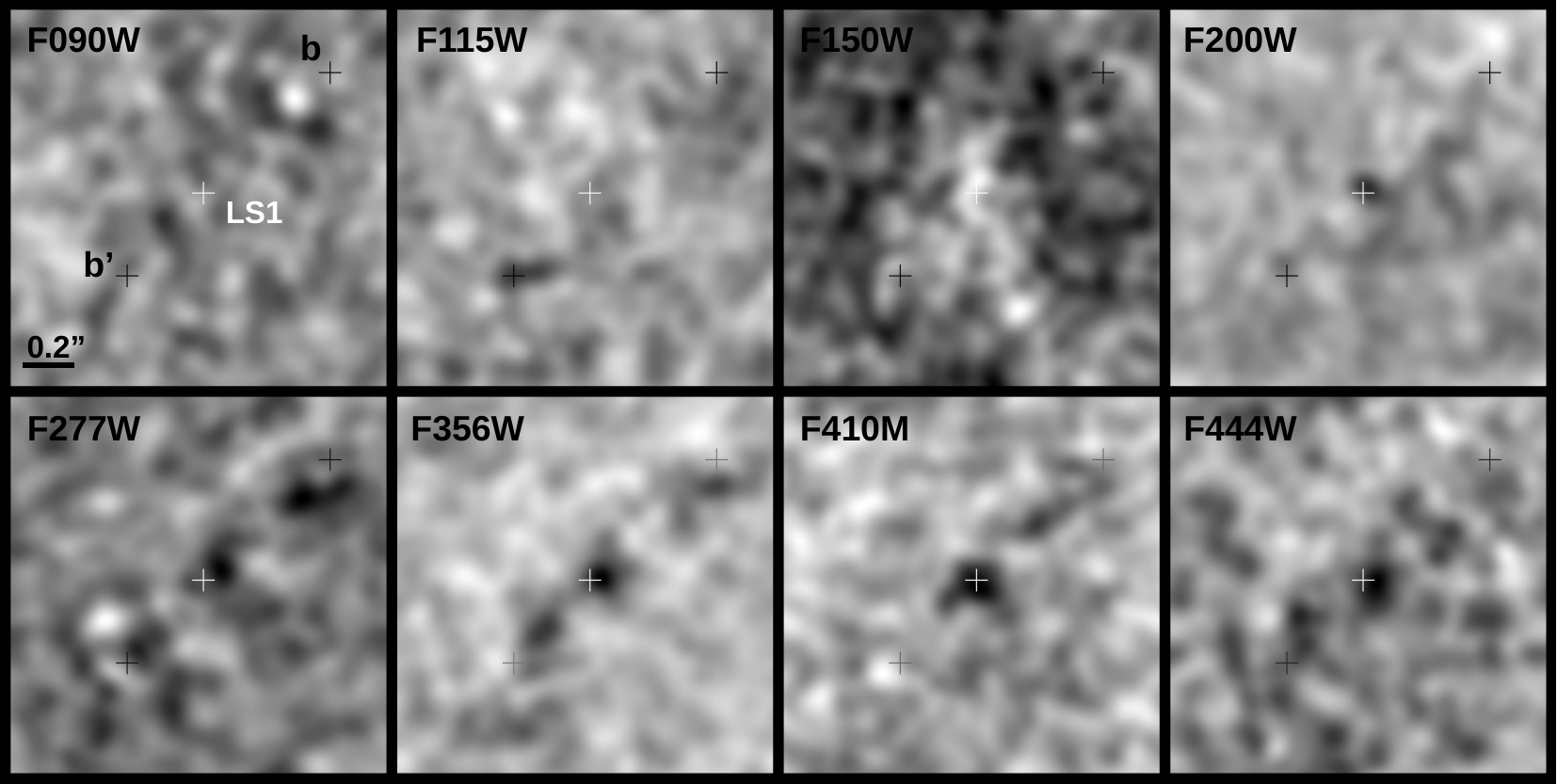}     
      \caption{Difference images between first and last epochs. Each panel shows a region of $1\farcs5\times1\farcs5$ centered on LS1 and in filters as labeled in each panel. A source that brightened between the two epochs shows up as dark in the figure. White crosses mark the position of LS1, and black crosses mark the positions of knot  b and its equally bright counterimage b$'$.   All images have been smoothed with a Gaussian kernel with $\rm FWHM=0\farcs09$ to increase contrast.  
         }
         \label{Fig_StackedDiffLS1}
\end{figure}

Difference imaging shown in Figure~\ref{Fig_StackedDiffLS1} confirms that the  variability is coming from LS1 and not some unrelated source. While some filters such as F277W,\, F356W, and F444W show a small offset between the variability peak and the position of LS1, this can be understood as a PSF effect. The NIRCam PSFs are asymmetric, and epochs 1 and 3 were taken at position angles (PA) differing by 42$^\circ$. PA differences have the biggest effect when the source is variable, as discussed further in Appendix~\ref{sec_VSVPA}.



\section{The likely counterimage of LS1}\label{sec_LS1prime}
The final piece of evidence in favour of LS1 being a lensed star would come from the detection of its counterimage, LS1$'$, on the other side of the critical curve. In a classic lensing scenario, where micro- or milli-lensing is not perturbing either of the images, we would see both images of the lensed star. Seeing an image pair would favor the lensing scenario over the projection-effect scenario because an object in the foreground would not be multiply lensed. As mentioned above, no LS1$'$ is directly detected in the images, but the bright arc limits the sensitivity. To subtract the arc, we created difference images in every possible wavelength pair after matching the image resolutions and in some cases normalizing the images. The results of all possible differences are shown in Figures~\ref{Fig_CornerPlot1},  \ref{Fig_CornerPlot2}, and~\ref{Fig_CornerPlot3}. 
Appendix~\ref{sec_CornerPlot} gives details.

\begin{figure}
\begin{center}
     \includegraphics[width=8cm]{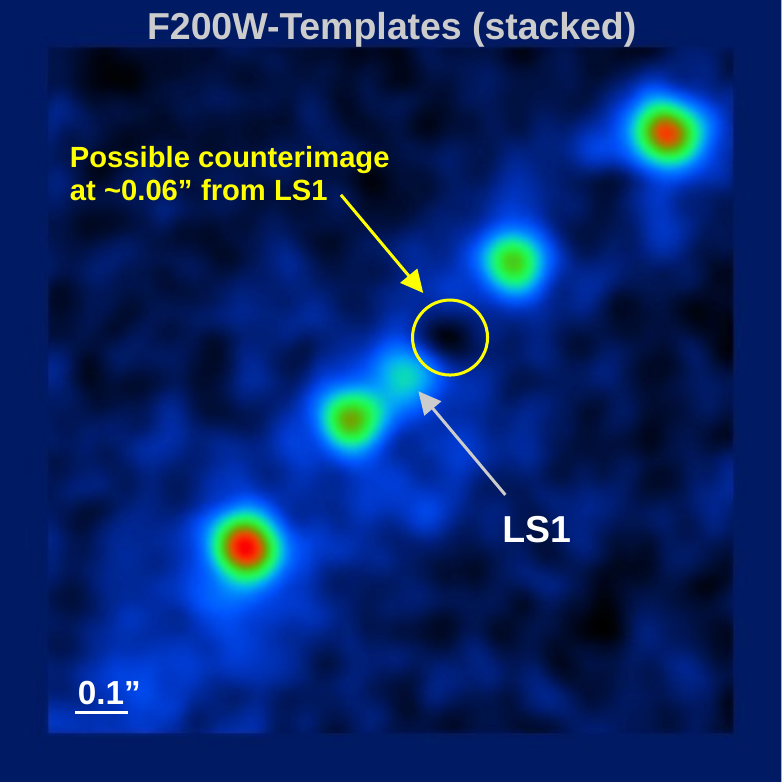}     
      \caption{Stacked difference $\sum_{i=1}^{i=3}(\mathrm{F200W}-\alpha_i\times \mathrm{FnnnW_i})$, where $\alpha_i$ was chosen to minimize the contribution from the arc to the difference, and $\mathrm{FnnnW}_i$ are all filters with wavelengths below 2\,\micron, that is F090W, F115W, and F150W. These were degraded to the resolution of F200W using the star-derived PSF in Appendix~\ref{sec_CornerPlot}. The individual differences are shown in column~3 of Figure~\ref{Fig_CornerPlot2}. The stacked image has been smoothed with a Gaussian of $\rm FWHM=0\farcs09$ to increase the contrast. The position of LS1 is marked with a white arrow. The position of the possible counterimage LS1$'$, $\approx$0\farcs1 from LS1, is marked with a yellow circle. }
         \label{Fig_Counterimage}
    \end{center}
\end{figure}

Most of the difference images show no evidence for additional point sources, but there are some differences in  $F200W-\alpha*F090W$ in Figures~\ref{Fig_CornerPlot2} and~\ref{Fig_CornerPlot3}. For this difference and also in  other $\rm F150W-F090W$ and $\rm F150W-F115W$, there is a significant negative fluctuation near LS1 (Figure~\ref{Fig_Counterimage}) close to the expected position of LS1$'$ if the critical curve is at the position of the dashed line in Figure~\ref{Fig_Star}. The significance of this negative fluctuation is ${\approx}4\sigma$, so we cannot rule out its being an unfortunate noise fluctuation, but the fact that similar (but less prominent) fluctuations can be observed in other differences suggests that this is possibly a real source. The difference shown in Figure~\ref{Fig_Counterimage} was obtained after adding all three differences shown in Figure~\ref{Fig_CornerPlot2} in the column labelled F200W. 

\begin{figure*}
\begin{center}
     \includegraphics[width=19cm]{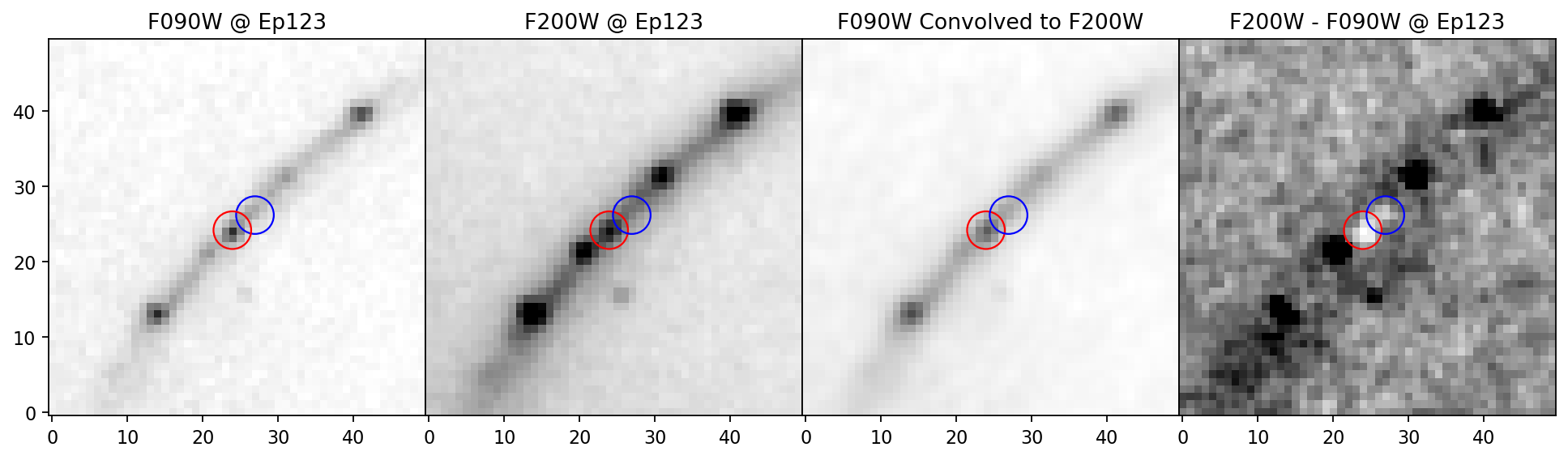}     
      \caption{An independent result confirming the likely counterimage. From left to right, the first and second panels show the (negative) original images of the arc in F090W and F200W; the third panel shows the convolved image from F090W to F200W using WebbPSF models; the fourth image shows $\rm F200W-1.87\times F090W$ to subtract off the local background in the neighbor of LS1. In all panels, red and blue circles show the LS1 and LS1$'$ positions, respectively. }
         \label{Fig_CounterimageTom}
    \end{center}
\end{figure*}

Seeing a negative signal, which for simplicity we refer to as LS1$'$, in the difference image implies a color difference from LS1, which is seen as a positive source.  This is only possible for a counterimage of LS1 if LS1 is a composite object, for instance a binary with a blue and a red component. The negative signal implies that in LS1$'$, the blue component is being magnified more than the red component. This is possible if microlensing affects the blue component more than the red one, which is possible for wide binaries or for a small group of stars with two dominant supergiants, one red and one blue. 
Given the relatively low significance of LS1$'$, we can not be certain this is the counterimage of LS1. Additional monitoring is needed in order to confirm or reject this hypothesis. There is so far no significant time variability at the LS1$'$ position between the different epochs with JWST data, but if microlensing is the cause of the chromatic effect, some variability should be observed in future observations at this position. 


Figure~\ref{Fig_CounterimageTom} shows a different difference-image search for LS1$'$. Here the F090W was PSF-matched to the F200W one using a convolution kernel generated with WebbPSF models. The local background near LS1 is ${\sim}1.87\times$ higher in the F200W image than in F090W, and  F090W was normalized by this factor and subtracted from F200W to minimize the local background in the neighborhood of LS1 and LS1$'$. The difference shows a clear negative signal $\approx$0\farcs1 from LS1.  There are even hints of a source at that location in the unsubtracted images.  LS1 and LS1$'$ both show negative signals, meaning that they are bluer than the local background. PSF photometry  using the WebbPSF model \citep[details described by][]{Haojing2023} gives flux densities {\em in the difference image} of $6.30\pm2.56$ and $3.64\pm2.56$\,nJy for LS1 and LS1$'$, respectively. This result is consistent with the one shown in Figure~\ref{Fig_Counterimage}. The   difference between the two results, a positive residue in LS1 and a negative one in LS1$'$ in the first and a negative residue for both LS1 and LS1$'$ in the second, could in part result from  the different PSFs (star model in the first case and WebbPSF in the second one). More likely, though, is the different normalization with LS1$'$ being compared to LS1 in the first case and to the local background in the second.

In summary, LS1$'$ is likely the counterimage of LS1, which would definitely confirm the strongly lensed nature of Mothra. In this case,  the critical curve must pass very close to the middle point between LS1 and LS1$'$. If so, the separation between LS1 and the critical curve is $D\approx 0\farcs055$ instead of $D\approx 0\farcs07$. This  would raise the macromodel magnification for LS1 and LS1$'$ to $\mu=62''/0\farcs055=1127$ and a physical separation for the binary of less than 0.75\,pc.

\section{Millilensing interpretation of LS1}\label{sec_millilens}

\begin{figure*} 
   \includegraphics[width=18cm]{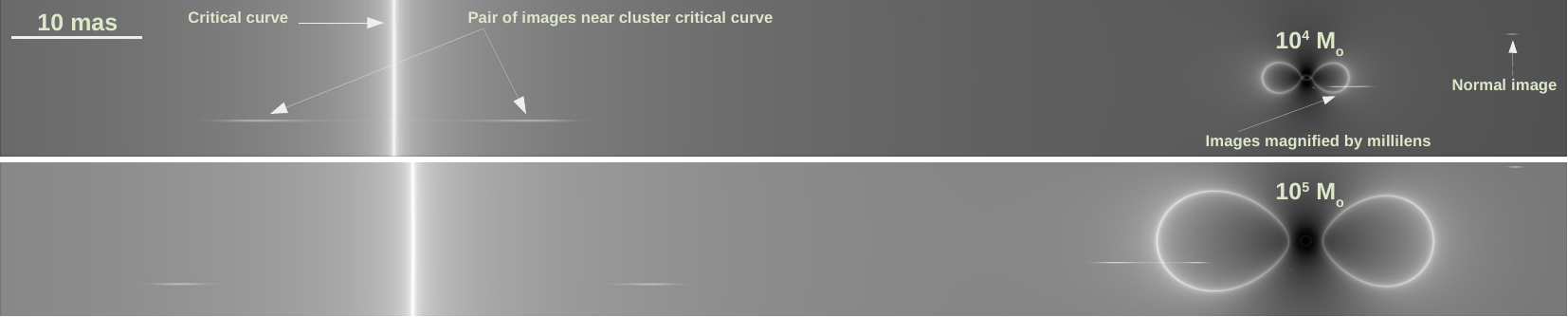}
      \caption{Millilens models near the critical curve. Grey scale shows the logarithm of the magnification at each location with the critical curve from the cluster being the white vertical line and the critical curves from the millilenses being white ovals. Each panel has a millilens 0\farcs07 (about twice the resolving power of JWST) to the right of the macrolens critical curve. The top panel is for a millilens with $10^4$\,\Msun, and the bottom panel is for a millilens ten times more massive.
      (The critical curve in the bottom panel moved towards the right owing to the contribution of the millilens to the total convergence.)
      Three sources, modeled as Gaussians with $\sigma=2\,\mmas=0.0168$\,pc and marked with arrows, are being lensed. 
      The source labeled ``normal image'' is far from the critical curve and magnified by the cluster macrolens with only a small contribution from the millilens. The source labeled ``Images magnified by millilens'' is magnified by the millilens. The source labeled ``Pair of images near cluster critical curve'' is close to the caustic of the cluster and forms a pair of images.         }
         \label{Fig_MacroPlusMillilens_2D}
\end{figure*}

\begin{figure*} 
   \includegraphics[width=8.5cm]{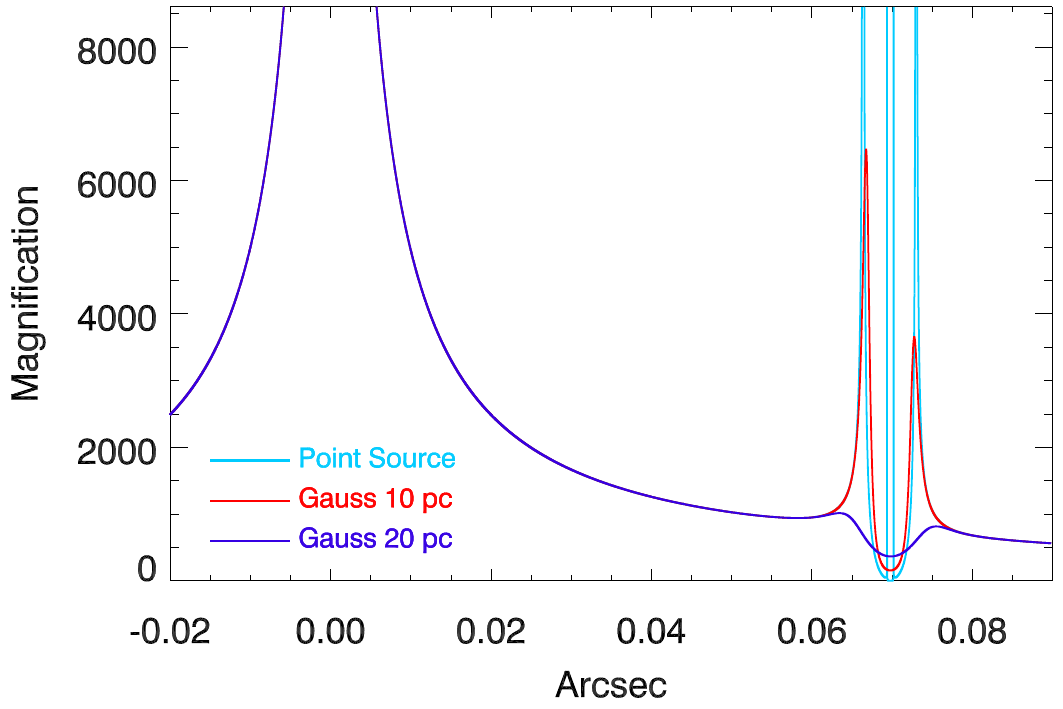}
   \includegraphics[width=8.5cm]{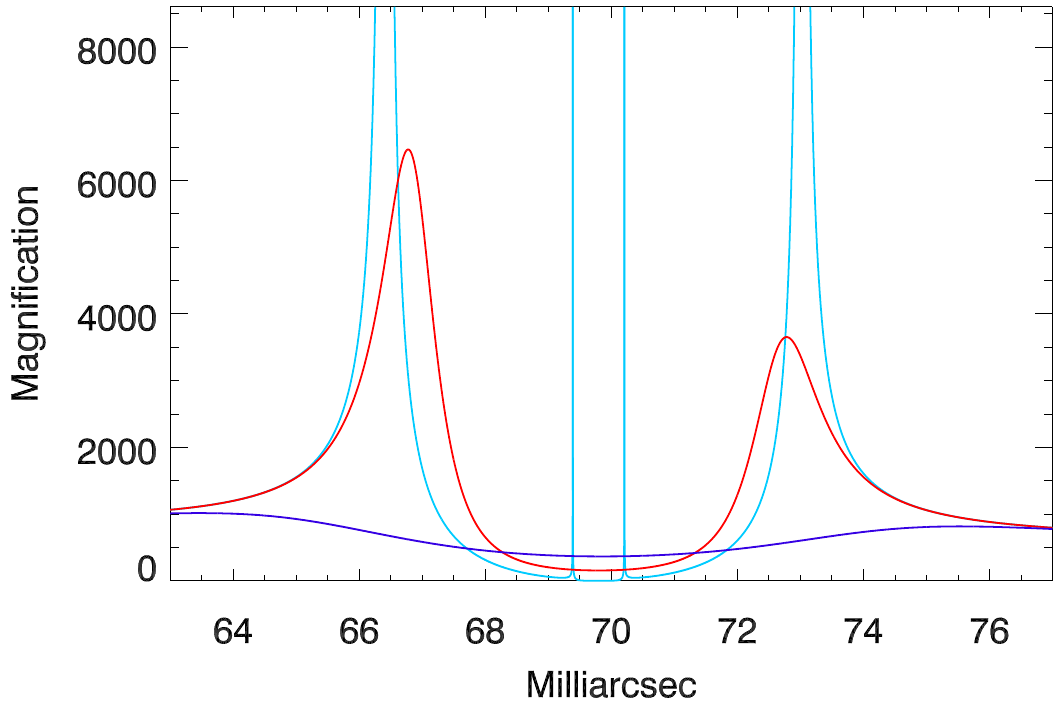}
      \caption{Simulated magnification across the arc in the lens plane. The critical curve is at the origin. A perturber with $10^4$\,\Msun\ is placed $\approx$0\farcs07 from the critical curve. Colors show three assumptions perturber  profiles: light blue for a point source, dark blue for a Gaussian profile with $\sigma=10$\,pc and red for a Gaussian profile with $\sigma = 20$\,pc.  Left panel shows a wide range, and the right panel is a magnified version around the perturber.
         }
         \label{Fig_Millilens1}
\end{figure*}

Having excluded the possibility that LS1 is a transient event, an unresolved double image on the critical curve, a projection effect (for instance one of the globular clusters found nearby) or a microlensing event, the only surviving hypotheses from Section~\ref{sec_LS1} is millilensing at the position of LS1. On this side of the critical curve, images have negative parity (saddle point) and have a relatively large probability of being demagnified with respect to the macromodel magnification \citep[e.g.,][]{Diego2018}. Because LS1 has been detectable for at least 8 years, while its counterimage LS1$'$ has remained undetected for the same period, the only possible explanation is that a substructure is boosting the magnification locally at the position of LS1 by at least an extra factor of 4 with respect to the macromodel value ($\mu_{macro} \approx 1000$). That makes the total magnification at the position of LS1 $\ge$4000. Microlenses can attain this only if the relative direction of motion and velocity of the background star is fine tunned. A more massive millilens has a larger caustic region which does not require any fine tunning. 


A millilens with $10^4<M_L<2\times10^7$\,\Msun\ can explain the large differential magnification between LS1 and a putative LS1$'$.  The source is observed only on the side with negative parity because only on this side can produce differential magnification $>$2\,mag. The more massive millilens also has a larger high-magnification region than a microlens, allowing the differential magnification to persist for more than eight years, independent of the relative direction of motion of Mothra. We therefore consider the millilens hypothesis the most likely explanation for the anomalous magnification of LS1.  At the low end of the mass range, $10^4\la M_L\la10^6$\,\Msun\ a possible source would be a small globular cluster that would not be detected in JWST images. On the high end, the millilens could be a massive globular cluster or galactic core remnant. Tidal stripping by multiple close passes by the BCG (projected distance $6\farcs3=33.6$\,kpc) could have made the object very compact. A central black hole would make the galaxy's mass even more compact. Remnant cores of very massive galaxies can, however, be ruled out because such a galaxy would have a central SMBH mass exceeding the upper limit derived below.

\subsection{Minimum mass of the millilens}\label{sec_minmass}
To study the millilens possibility in more detail, we constructed a set of lens models with $M_L$ between $10^4$\,\Msun\ and $10^7$\,\Msun.  Each model had a millilens forming a smaller critical curve $\approx 70$\,mas from the macrolens critical curve, and we simulated an area large enough to contain the source and the critical curve.  The pixel size was 50\,\mmas\ for the higher-mass lens and 1\,\mmas\ for the lower mass.  The latter scenario considered three mass distributions for the millilens: a point source, a Gaussian with $\sigma=10$\,pc, and a Gaussian with $\sigma=20$\,pc.  (These correspond respectively to an intermediate-mass black hole or core-heavy globular cluster, a loose globular cluster, and a small dwarf galaxy.)  The macromodel magnification was fixed to $\mu=800$ with radial component $\mu_r=3.3$ as predicted by the lens model. Microlenses from the intracluster medium were not included in the simulation, except for the most massive millilens (just for illustration purposes. Figure~\ref{Fig_MacroPlusMillilens1E7_2D} shows a case with microlenses).
Figure~\ref{Fig_MacroPlusMillilens_2D} shows the resulting magnifications near the critical curve for three sources at different distances from the critical curve, and for two different millilenses.  All three sources are stretched into lines (because $\mu_t\gg\mu_r$), but this effect is below JWST's resolution.  All three sources produce image pairs, one image on each side of the cluster critical curve, but the counterimages for the two sources most distant from the critical curve are outside the image boundary on the left.
The source near the caustic of the macromodel shows two images near the critical curve with $\mu>10000$. The source  magnified by the millilens has $\mu\sim4000$, but its counterimage has only the macrolens magnification of $\approx$800.  The source farthest from the critical curve is magnified mostly by the cluster and has $\mu \sim 1000$. 
For the  $M_L=10^5$\,\Msun\ case, the critical curves increase in radius by ${\approx}\sqrt10$, and the probability of magnifying a background source by factors of several thousand increases accordingly.


Figure~\ref{Fig_Millilens1} shows the model magnifications computed in the image plane. The point source produces sharp magnification peaks and dips. The smoother Gaussian profiles can be subcritical, that is, not producing critical curves around the millilens. Criticality can be obtained for smaller values of $\sigma$. For instance for $\sigma=1$\,pc, the deflection field can produce real critical curves around the deflector. Despite being subcritical, a deflector with mass $M=10^4$\,\Msun\ and  $\sigma \lesssim 10$\,pc can produce magnification factors large enough and lasting $>10$~years independent of the direction of motion of the background source provided it is close enough to the region of maximum magnification.

\begin{figure} 
   \includegraphics[width=8.5cm]{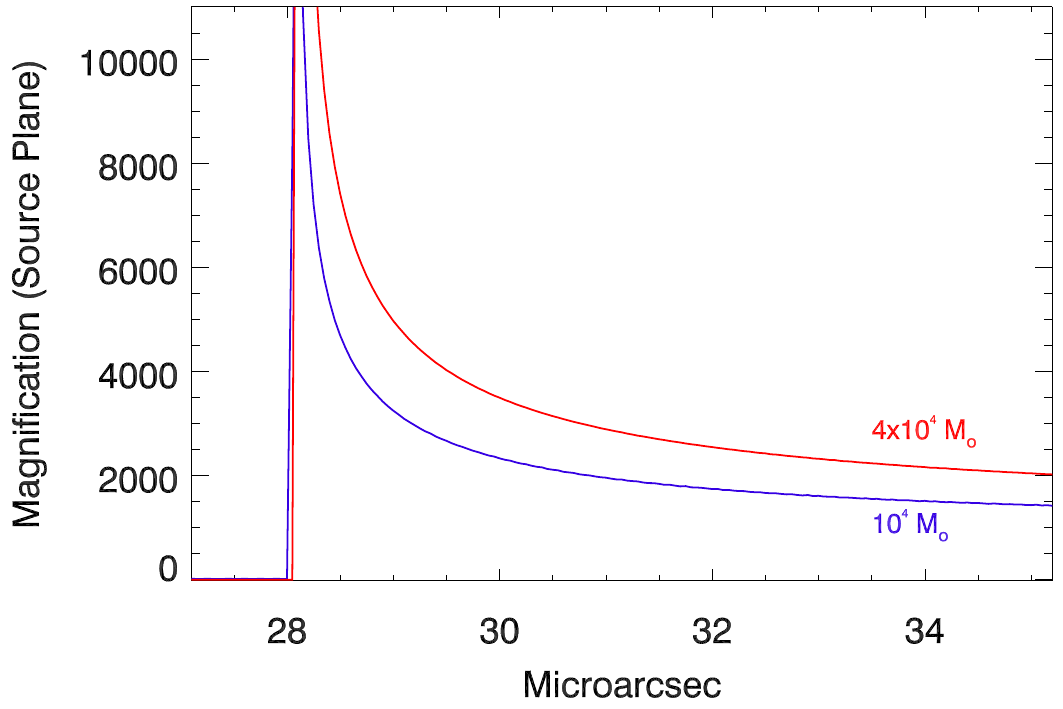}
      \caption{Magnification in the source plane in one of the two millicaustics. Curves are for $M_L=10^4$\,\Msun\ (lower) and $4\times10^4$\,\Msun\ (upper). The origin is at the projected position of the millilens. Both curves are for a Gaussian millilens profile with $\sigma = 2$\,pc. 
         }
         \label{Fig_Millilens3}
\end{figure}

Figure~\ref{Fig_Millilens3} shows a high-resolution view of the caustic in the source plane.  At distances $\la$28\,\mmas\ from the origin, $\mu \approx 10$, much smaller than the macromodel $\mu=800$.   Just outside this region, the magnification can be very high, and there is a range $\approx$1\,\mmas\ wide where  $\mu>4000$. A star in this region would be brighter by $>$1.75 magnitudes compared with stars magnified only by the macromodel. At relative speed of 1000\,km\,s$^{-1}$, a star moving directly toward the millicaustic takes $\approx$8\,years to cross this distance. If it is moving more nearly parallel to the millicaustic, it would take longer to leave the region of high magnification. It would also take longer if the transverse velocity is lower. If the mass of the millilens is larger, the thickness of the region parallel to the microcaustic with magnification greater than 4000 increases linearly with the mass of the millilens \cite{Palencia2023}. This results in an increase of the thickness of the caustic region with $\mu>4000$ by a factor $\approx 2$.
Considering all these factors, millilens masses $>$$10^4$\,\Msun\ can easily provide $\mu> 4000$ for $\ge$8\,years.

\subsection{Maximum mass of the millilens}\label{sec_maxmass}

A simple upper limit on the millilens mass can be established from the lack of obvious lensing magnification on the nearby knot c$'$. If the Einstein radius of the millilens were ${\ga}1/3$ of the distance between LS1 and c$'$, c$'$ would be magnified by much more than c. This requires $\Theta_E<0\farcs043$. 
$\Theta_E\approx 0\farcs003$ for the $10^4$\,\Msun\ millilens,
and the Einstein radius scales as the square root of the mass. This requires $M_L\la 2\times10^7$\,\Msun\ in order to magnify LS1 but not c$'$. A millilens with this mass would have a large region with $\mu> 4000$, but, as shown in section~\ref{sec_globclust}, a stellar population with this much mass would be luminous enough to have as much flux as LS1.

\begin{figure*} 
   \includegraphics[width=18cm]{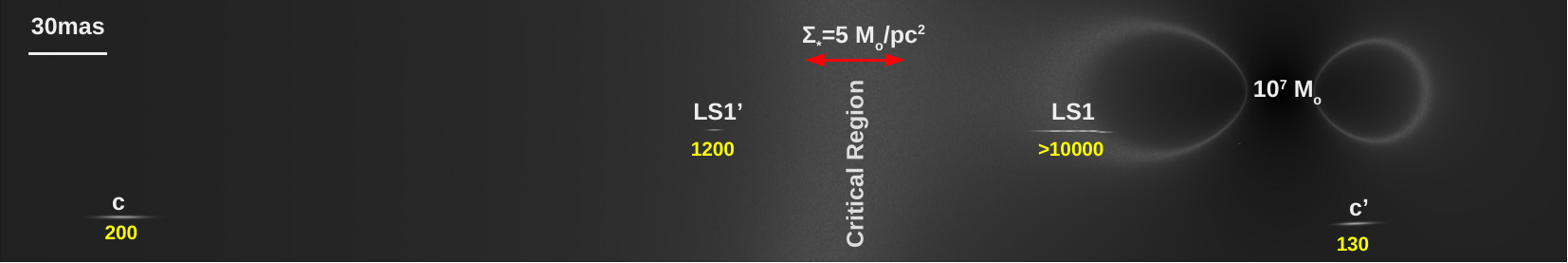}
      \caption{Millilens model with mass $10^7$\,\Msun\ near the critical curve and magnifying LS1. The millilens is  farther from the critical curve than in Figure~\ref{Fig_MacroPlusMillilens_2D} but still forms critical curves at the position of LS1. This simulation includes both LS1 and i LS1$'$ as well as the knot image pair b and b$'$. The geometry of the simulation mimics the observations, with a separation between b and b$'$ of 0\farcs45 and a tangential separation (i.e., in the vertical direction) between LS1 and knot b$'$ of 30\,mas. The numbers in yellow indicate the approximate magnification at the position of the corresponding counterimage. In this configuration, LS1$'$ is $\approx 2.5$\,magn fainter than LS1 and therefore unobserved.  For this mass, the effect on knots b and b$'$ is noticeable with the magnification of b$'$ affected by the millilens. Stars in the intracluster medium with a surface mass density of 5\,\Msun\,pc$^{-2}$ blur the critical curves and reshuffle large  magnifications away from the critical curve. The red double-arrow segment shows the width of the saturation region as predicted by \citet[][their Eq.~15]{Diego2018}.  
         }
         \label{Fig_MacroPlusMillilens1E7_2D}
\end{figure*}

A more detailed model sets a tighter limit on the maximum mass of the millilens.  
Figure~\ref{Fig_MacroPlusMillilens1E7_2D} shows the case of a massive millilens with mass equal to $10^7$\,\Msun. The pixel scale for this simulation is 50\,\mmas. The millilens was placed to put its critical curves 70\,mas from the critical curve of the cluster and avoid merging the cluster and millilens critical curves.  
To simulate the background sources, we again adopted Gaussian profiles with LS1's source (Mothra) having $\sigma=0.04$\,pc. Stars would be much smaller than this and better approximated by a disc, but the Gaussian model suffices for illustration. The model accounts for the effect of the millilens on the neighboring knots c and c$'$. The source of these knots was modeled as a Gaussian  with $\sigma=0.85$\,pc, mimicking a  compact star-forming region or globular cluster. The positions in the source plane were adjusted to make all lensed images resemble the observation. The result is shown in Figure~\ref{Fig_MacroPlusMillilens1E7_2D}. The ratio of magnifications between LS1 and LS1$'$ is $>$5, as required in order to observe LS1 but not LS1$'$. 

For illustration purposes, the model includes the effect of microlenses, ignored so far. For the microlenses, we used a simple approximation\footnote{A real simulation over an area of 0\farcs6 that resolves stellar microlenses would require a pixel size smaller than 1\,\mmas. That would require ${>}10^9$~pixels, which is computationally demanding. The simple approximation used here is accurate enough for present purposes.} that the deflection field is a Gaussian random field \citep{Dai2021} with dispersion directly proportional to the surface mass density of microlenses, $\Sigma_*$. Real $N$-body simulations in smaller areas have shown that $\sigma \approx 0.036\Sigma_*/($\Msun pc$^{-2})$\,\mmas, and 
the model used a conservative $\Sigma_* = 5$\Msun pc$^{-2}$. We included microlenses by adding a random Gaussian field to $\alpha_x$ and $\alpha_y$ with dispersion $\sigma=0.036\Sigma_* = 0.18$\,\mmas. 
For this value of the surface mass density, $A=62\arcsec$ and $\mu_r=3.3$, and the width of the saturation region, i.e., the region where microlenses fully disturb the critical curve region, is $\Delta\Theta=4.2\times 10^{-4}\times \Sigma_* \times A / \mu_r=0\farcs04$ \citep[][their Eq.~15]{Diego2018}. Our simple simulation (Figure~\ref{Fig_MacroPlusMillilens1E7_2D}) reproduces this width. Even for point sources as small as stars, the magnification within the saturation region typically cannot exceed $\sim$100\,000. 
In the presence of microlenses, the sharp critical curves are substituted by a corrugated network of less powerful microcritical curves. Extreme magnification factors of order one million or more for background stellar objects, possible when microlenses are not present, are prohibited when microlenses are included in the calculation. To compensate, regions that are farther from the critical curve, where without microlenses the macromodel magnification would be O(1000), can momentarily magnify a background star by O(10\,000). The time-integrated flux of a background star crossing the entire network of microcaustics (taking hundreds or even thousands of years) is the same whether microlenses are present or not, as required by flux conservation.

\begin{figure} 
   \includegraphics[width=8.5cm]{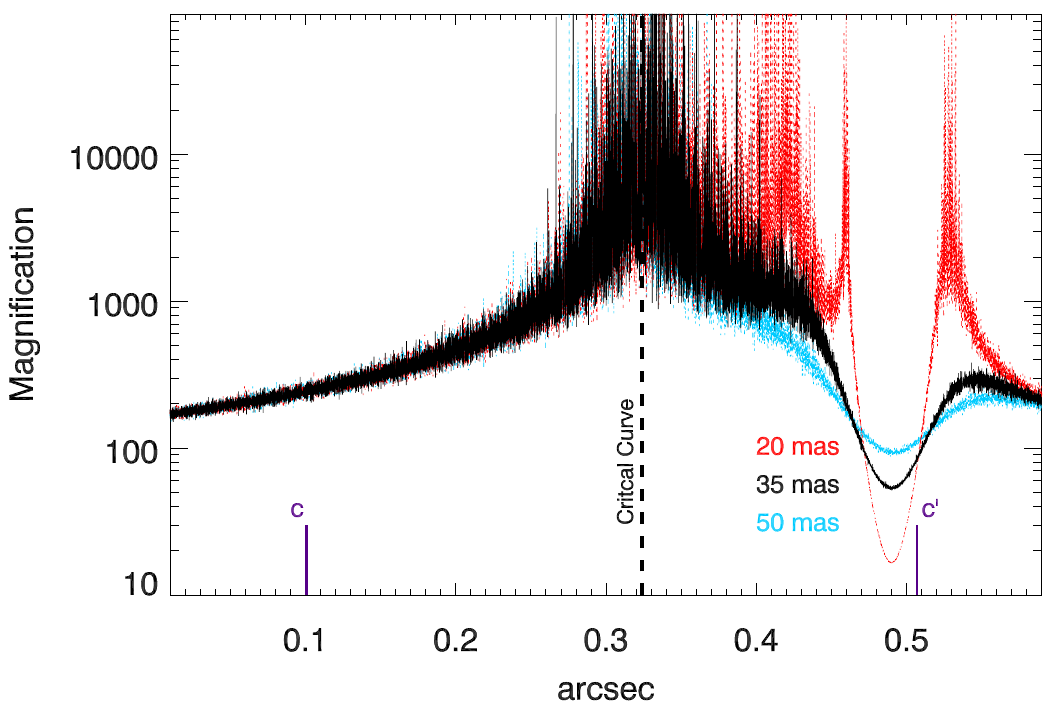}
      \caption{Magnification of knots c and c$'$ for the model shown in Figure~\ref{Fig_MacroPlusMillilens1E7_2D}. The y-axis shows the magnification along lines perpendicular to the critical curve.
      The small-scale fluctuations are due to microlenses.  
      The position of the critical curve is shown as a vertical dashed line and the positions of c and c$'$ by vertical solid lines. The black curve shows the magnification along a horizontal line in Figure~\ref{Fig_MacroPlusMillilens1E7_2D} crossing 35\,mas south of the millilens.  The other two lines are at 20\,mas and 50\,mas as labeled. The three lines cover the range of estimated distances between c$'$ and the possible millilens.
         }
         \label{Fig_MillilensMagnif}
\end{figure}

\begin{figure*} 
   \includegraphics[width=18cm]{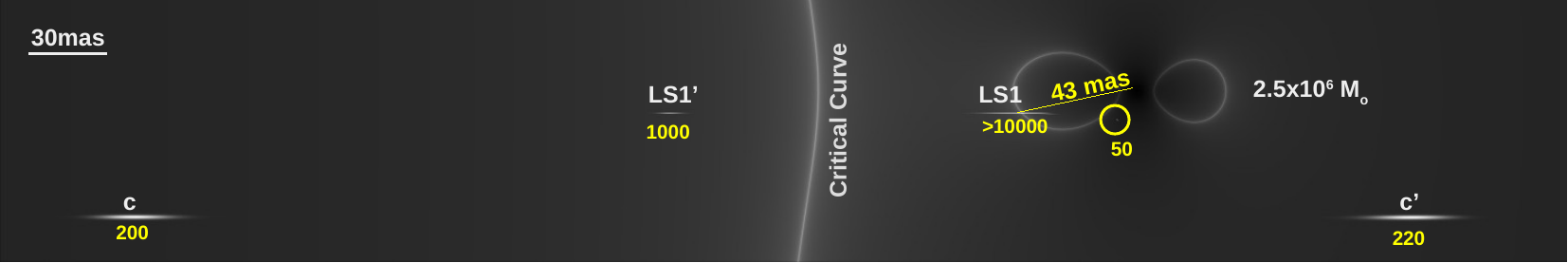}
      \caption{Model for $M_L=2.5\times10^6$\,\Msun, the maximum mass allowed for a millilens. Microlenses will be present and distort the critical curves as in  Figure~\ref{Fig_MacroPlusMillilens1E7_2D}, but their effect is omitted for clarity. 
      The magnification for each image is indicated in yellow. The millilens is  43\,mas from the magnified image LS1 as shown by the labeled scale bar. A third predicted image of Mothra with magnification $\approx$50 (therefore undetectable) is marked by a yellow circle.
         }
         \label{Fig_Millilens_2p5E6}
\end{figure*}

The situation represented in Figure~\ref{Fig_MacroPlusMillilens1E7_2D} corresponds to an extreme situation where LS1 is the result of the merging of two microimages, each with magnification $\sim$10\,000. The result is $\ga$3\,mag difference between LS1 and LS1$'$. If the two microimages are close enough to each other, the pair appears as an unresolved single source. If the source of LS1 moves farther away from the caustic of the millilens, the magnification is reduced, but the separation between the pair of microimages responsible for LS1 increases. For magnification factors $\approx$10\,000, the separation between microimages is comparable to the pixel size in NIRCam images, which is approximately the resolving power of the telescope.  Smaller magnification factors (i.e., larger separations) would make the source appear resolved, in conflict with the observations.

The Figure~\ref{Fig_MacroPlusMillilens1E7_2D} model predicts magnification factors for knots c and c$'$ between 130 and 200. The observed flux ratio is $c/c' \approx 1$, while the magnification ratio derived from the geometric distances between knots b--c and b$'$--c$'$ is $(b-c)/(b'-c')\approx 1.25$. This magnification ratio is given by the macromodel magnification and is basically unaffected by microlenses or millilenses with masses similar to those considered. 
However, the Figure~\ref{Fig_MacroPlusMillilens1E7_2D} simulation predicts $\mu_c/\mu_{c'}\approx 1.4$, inconsistent with the observations. 
This is better illustrated in Figure~\ref{Fig_MillilensMagnif}, which shows a 1-dimensional plot of the magnification pattern of the model. The magnification of c$'$ is clearly smaller than the magnification of c in all cases because c$'$ falls within the demagnification region of the millilens. 
Thus the millilens increases the flux ratio between c and c$'$, in contradiction with observations. Eliminating the discrepancy requires a smaller millilens mass, which will make c$'$ fall in the magnification region and bring $\mu_{\rm c}/\mu_{\rm c'}$ closer to the observed flux ratio of 1. 
This sets an upper limit ${\ll}10^7$\,\Msun\ on the mass of the millilens.
This simulation has LS1 in the ``lower'' branch of the oval critical curves around the millilens. Because the tangential separation between LS1 and c$'$ is fixed, putting LS1 in the ``upper'' branch would move  c$'$ closer to the millilens, and the magnification would be even larger.
This would require an even smaller mass for the millilens not to violate observational constraints.

Even with a mass of $10^7$\,\Msun\ for the millilens, it is difficult to simultaneously explain LS1, its lack of counterimage above the detection limit, and the separation and relative flux of c and c$'$.  In addition, a millilens {\em stellar} mass ${>}10^7$\,\Msun\ would be directly visible in the JWST images (Section~\ref{sec_globclust}). The distance from the millilens to its critical curves is $>$60\,mas (Figure~\ref{Fig_MacroPlusMillilens1E7_2D}), and the millilens--LS1 pair would be resolved, contradicting observations.  The JWST resolution limit is nearer 30\,mas, and the limit on the size of the critical curves can be scaled accordingly. The mass goes as the square of the size of the critical curve, and therefore the mass limit goes down by a factor of four. If stellar system with a mass of $2.5\times 10^6$\,\Msun\ would be detectable by JWST, we can rule out masses above this one on the grounds of the unresolved nature of LS1. Hence we adopt an upper limit for the mass of the millilens of $2.5\times10^6$\,\Msun. This is about the minimum stellar mass that would be detectable by JWST at M0416's distance (see section~\ref{sec_LS1}).

One final issue to consider is that some or all of the millilens mass could be in the form of a black hole. Black holes are commonly found at the centres of ultra-compact-dwarfs (UCDs) orbiting around massive galaxies, such as BCGs. If the millilens contains a SMBH, by the same arguments discussed above, this should have a mass smaller than $2.5 \times10^6$\,\Msun. 

A model for a millilens with  $M_L=2.5\times10^6$\,\Msun\ (and Gaussian profile as before) is shown in Figure~\ref{Fig_Millilens_2p5E6}. The spatial configuration of sources and millilens reproduces the observations. The magnification ratio between LS1 and LS1$'$ is $>$4 as required. A third counterimage is predicted closer to the millilens but with much smaller magnification and therefore undetectable. This third counterimage would be even smaller with a more cuspy profile but with the same total mass. The model flux ratio between c and c$'$ is 0.9, close enough to 1.0 to agree with observations. Finally the separation between the millilens and LS1 is comparable to the resolving power of JWST, so even if the millilens is luminous enough to be observable by JWST, the LS1+millilens pair would still appear unresolved. The model is thus fully compatible with the observations.

\section{Discussion}\label{sec_discussion}
Summarizing our main results, LS1 is detected with F200W apparent magnitude $\approx$27.8\,AB at $\approx$0\farcs05 from the critical curve. The macromodel magnification at this position is predicted to be O(1000). While the F200W and shorter-wavelength images show hints of LS1$'$, the counterimage of LS1, at $\approx$0\farcs1 from LS1, LS1$'$ has an apparent magnitude $>$29.5. This translates into a differential magnification $>$4 between LS1 and LS1$'$.  Like LS1, LS1$'$ should have a macromodel magnification of O(1000). Hence, the magnification of LS1 must be $\ga$4000. A millilens placed along the line of sight to LS1 provides the needed extra boost in magnification making the net magnification of LS1 $\ga$10\,000. 

The lensed background source Mothra is likely composed of at least two supergiant stars, one with $T\approx5000$\,K and the other one with temperature $T \approx 14000$\,K. 
Adopting an observed luminosity $\log_{10}(\mu L_{\rm bol}/L_{\odot})=8.4$ for the red SG, and assuming the magnification factor is 5000, this implies an intrinsic bolometric luminosity $L_{\rm bol} > 5\times10^4$\,\Lsun. The bolometric luminosity of the blue SG is approximately 2.5 times larger. 

The only plausible scenario for the lens is for it to have a mass at least $10^4$\,\Msun\ in order to create a millicaustic with a region around the caustic having magnification factors at least 4000 ($2.5\log_{10}(3500/825)\approx 1.5$\,mag boost in relation to LS1$'$) and thickness at least 1\,\mmas\ (distance travelled by a source moving at 1000\,km\,s$^{-1}$)
so LS1 can appear magnified by $\mu>4000$ during at least 8 years without requiring it to move exactly parallel to the millicaustic. On the other hand, the mass of the millines needs to be less than  $2.5\times10^6$\,\Msun\ in order to not introduce anomalous flux ratios in the counterimages c and c$'$, and/or be directly detectable by JWST and produce a resolved image of the pair LS1+millilens. A mass of ${\sim} 10^5$\,\Msun\ offers a good compromise because it can easily accommodate both constraints and produce negligible effects on the c/c$'$ flux ratio.

\subsection{Time variability: Intrinsic or microlensing?}
The observed time variability could arise from intrinsic variability of the redder star in the Mothra binary.
An alternative possibility is that the red SG star is  moving close to one of many microcaustics, while the blue SG remains too far away to undergo any significant change in flux. 
If the two stars are separated by a relatively large distance, $d > 0.01$\,pc (consistent with the size constraint, $d<1$\,pc derived from the fact that LS1 is unresolved at magnifications factor O(1000)), this translates to an angular separation $>$1\,\mmas. This is large enough to allow a 1\,\Msun\ microlens (Figure~\ref{Fig_Microlens1}) to temporarily magnify  the red component but not the blue one.
In this case, the red SG  flux change $\Delta F\propto 1/{|t-t_o|}^{-0.5}$, where $t_o$ is the time the microcaustic was crossed. This interpretation is, however, challenged by the small flux change between Epc and Ep3. This constancy during a caustic crossing can be attained if the red SG portion of Mothra is moving with a relatively small velocity with respect to the microcaustic, and the radius of the red SG is large.   A star with radius 300\,R$_{\odot}$ moving with relative velocity 100\,km,s$^{-1}$ with respect to a microcaustic moves one stellar radius ($\approx$1 nanoarcsec) in one month, the time separation between Epc and Ep3. While possible, this scenario requires fine tuning between the three planes (observer, cluster, and source) or the direction of motion of the star (for instance, close to parallel to the microcaustic) in order to produce such small relative velocities. 
This makes intrinsic variability of the red component of Mothra a more likely hypothesis.  
This is not surprising as red SGs are known to be variable. Intrinsic variability of the red component opens the possibility to estimate its intrinsic luminosity (and hence derive the magnification) based on its observed periodicity \citep[e.g.,[]{Soszynski2007,Yang2012}. Such a feat requires long-term monitoring of this arc, but repeated observations of M0416 are well motivated because lenses a wealth of background sources, some of which are expected to be intrinsically variable red SG stars \citep{Diego2023c,Haojing2023}.



\subsection{The millilens in cold dark matter models}
The mass of the millilens is constrained to be between $10^4$\,\Msun\ and $2.5\times 10^6$\,\Msun. A natural possibility is to consider one of the globular clusters (marked in Figure~\ref{Fig_Star}) near the lensed galaxy. The observed globular clusters are more massive than the millilens, but less massive clusters would be too faint to observe and could exist.
If an intermediate-mass black hole \citep{Seth2014} contributes part of the millilens mass, the luminous mass would be correspondingly smaller and the clusters even fainter.
This picture is  consistent with the standard CDM model that predicts a wealth of dark matter halos in this mass range. These halos would contain also baryons that can cool down more efficiently than dark matter and form compact nuclei at the center of the halos. As the low-mass halos fall into the deepest regions in the cluster potential well, the outer parts of the halo get stripped away by tidal forces \citep{Chilingarian2023}, leaving a compact core. This mechanism works in a wide range of masses and has been invoked to explain for instance the presence of ultracompact dwarf galaxies in galaxy clusters \citep{Pfeffer2013,Mihos2015}. 

Below we consider two alternative models of dark matter, that can predict a different distribution of matter below the kpc scale. 

\subsection{Implications for the warm dark matter model}
Assuming the millilens has significant contributions from DM we can set constraints on models of warm DM (or WDM)\null. If the free-streaming length of the DM particle is too large, DM halos do not form below this scale, and the resulting halo mass function exhibits a cutoff at a characteristic mass \citep{Bond1983,Benson2013}. 
In WDM, the half-mode mass of the halo mass function can be related to the DM particle mass by the relation
\citep{Schneider2013,Gilman2020}
\begin{equation}
    m_{\rm hm}=3\times10^8\left(\frac{m_{\rm DM}}{3.3 \, {\rm keV}}\right)^{-10/3} \, \hbox{\Msun}.
\end{equation}
If we assume the half-mode mass is $2.5\times10^6$\,\Msun\ (the WDM mass function is suppressed by a factor 1/2 at the scale of the half-mode mass), this results in a minimum mass for the DM of 14\,keV. This constraint pushes the limit of previous results by a factor $\approx$2 \citep{Irsic2017,Hsueh2020,Gilman2020}. This result does not take into account that a halo of dark matter orbiting near the BCG would have a fraction of its mass stripped away by tidal forces. Hence this constraint should be considered with caution because the original halo would have been more massive (hence lowering the minimum mass for the DM particle). Results based on simulations suggests that up to 80\% of the dark matter mass can get stripped away from infalling satellites into massive clusters \citep{Niemiec2019}. If this much mass was lost, the mass of the DM halo before infall would have been 5 times larger, that is, $1.2\times10^7$\,\Msun\ and the minimum mass for the DM particle decreases to 8.7\,keV. 

\subsection{Implications for the fuzzy dark matter model}
Another interesting possibility is fuzzy dark matter (or FDM, also known as wave dark matter or axion-like-particle dark matter among other names). In this model, the dark matter can be described by a scalar field with an associated particle that is extremely light. Owing to its low mass, the associated wavelength of the particle is at the astrophysical scale. Relevant for our work is that in this model, the density field of dark matter fluctuates in length scales of the de Broglie wavelength of the DM particle ($\lambda = h/m_av$) on $\sim$Gyr timescales. These fluctuations are often referred to as granules, and they are characterized by their physical size and mass.  The granules can produce distortions along the critical curve, naturally predicting anomalous flux ratios between pairs of images \citep{Amruth2023,Powell2023}. The time variability of these granules is many orders of magnitude larger than the time spanning our observations and can be ignored.

A proper study of FDM requires constructing models with the constraints provided by the macrolens model. This is  well beyond the scope of this paper, but we can present a simple analysis. In the popular axion mass range of $m_a = 10^{-22}$\,eV, the de Broglie wavelength of the DM granules is approximately 0.3\,kpc \citep[for velocity $v=400$\,km\,s$^{-1}$][]{Laroche2022}. 
This is small enough to remain unresolved at the cluster distance, and these granules could act as millilenses. In the classic picture of FDM, the density fluctuations of the granules are of the same order of magnitude as the density itself \cite{Schive2014,Dalal2021}. In galaxy clusters at distances of 50--100\,kpc from the BCG (comparable to the distance from Mothra to the BCG), the dark matter density is typically 1--10\,$10^6$\,\Msun\,kpc$^{-3}$. Thus a DM granule could have the required mass and scale to act as the millilens discussed in Section~\ref{sec_millilens}. For a galaxy cluster as massive as M0416, there would be hundreds of granules with similar density fluctuations (half of them negative and half of them positive) projected along the line of sight. Their lensing distortions would partially compensate each other, so the net effect is expected to be smaller. Nevertheless, we expect random fluctuations equivalent to the contribution from a few granules along the line of sight. These fluctuations should be ubiquitous along the critical curves  \citep{Amruth2023} and leave imprints on other knots in the same arc such as c and c$'$ in the form of flux distortions. In fact, there is no evidence for significant flux distortions on c and c$'$. Larger axion masses would form granules with smaller de Broglie wavelength and smaller mass per granule, resulting in smaller lensing distortions. The mass of the granule scales as $\lambda^3$, and an axion mass 5 times larger results in granule masses around two orders of magnitude smaller, below the lower limit for the millilens ($M_{\rm min} \sim 10^4$\,\Msun). Therefore axion masses above ${\approx} 5\times 10^{-22}$\,eV cannot produce millilens-like fluctuations with the required mass to explain the  observations. Lighter axion masses can create granules that are more massive and give larger lensing distortions, but in this case, the DM granules extend over larger scales. These granules can be described to first order as Gaussian perturbations, and  Section~\ref{sec_millilens} showed how millilens Gaussian profiles with large $\sigma$ may become undercritical at the position of LS1 and unable to produce the required large magnification factors. For a particular case of an axion mass of $0.5\times 10^{-22}$\,eV, the mass of the granule would be ${\sim}10^7$\,\Msun. A Gaussian granule with this mass, placed at the right position to magnify Mothra to factors of several thousand, does not produce critical curves if $\sigma>225$\,pc. Masse  ${<}0.5\times 10^{-22}$\,eV are then in conflict with the de Broglie wavelength which would be $\lambda \approx 600$\,pc for this particular axion mass model. 
Therefore the mass of the axion cannot be much smaller than $10^{-22}$\,eV in order to produce extreme magnification at the position of LS1. As a conservative approximation, we consider $0.5\times10^{-22}$\,eV to be the lower limit for the axion mass consistent with the observations. 
Masses in the range $5\times 10^{-23}< m_a <5 \times 10^{-22}$\,eV are then (in principle) consistent with the lensing constraints discussed in this paper. As noted earlier, this conclusion is based on significant approximations, so it should be regarded with caution. 


\section{Conclusions}\label{sect_concl}
Mothra, a new kaiju (or monster) star at redshift $z=2.091$, is being magnified to extreme factors by the combined effect of a massive galaxy cluster and a millilens along the line of sight. 
The star is best described as a binary system with two supergiant stars, a hot one with temperature $T\approx 14000$\,K, and a cooler one with  $T\approx 5000$\,K. Mothra's flux changed at observed wavelengths $\ge$1.5\,\micron, which we attribute to intrinsic variability in the cooler star. 
A possible counterimage may exist (with low confidence) at the expected position on the opposite side of the lens-model critical curve, but it must have significantly lower magnification than the main image. Hypotheses to explain the anomalous magnification of Mothra and its counterimage include a transient event, a globular cluster or foreground object, a cluster-caustic crossing, microlensing, or a millilens. Among these, only millilensing offers a satisfactory explanation. In particular, the fact that Mothra was detected in 2014 as part of the HFF program is key to constraining the minimum mass of the millilens to ${>}10^4$\,\Msun.  At the opposite extreme, the lack of distortions on a nearby image pair (observed for the first time by JWST) and the fact that no  millilens is detectable in the JWST images set an upper limit on the mass of the millilens ${\le}2.5\times10^6$\,\Msun.

One possibility for the millilens is a small globular cluster, undetected in JWST images. Because this millilens is undetected, it could be dominated by dark matter. Such a millilens sets significant constraints on models of dark matter. Our findings are consistent with expectations from the cold dark matter model, but warm DM models where the DM particle is lighter than 8.7\,keV are excluded by the existence of such a millilens. Models of fuzzy dark matter could in principle reproduce the observations but only if the axion mass is  $5\times 10^{-23}< m_a <5 \times 10^{-22}$\,eV. 

\begin{acknowledgements}
J.M.D. acknowledges the support of project PGC2018-101814-B-100 (MCIU/AEI/MINECO/FEDER, UE) Ministerio de Ciencia, Investigaci\'on y Universidades. 
AZ, LJF and AKM acknowledge   support by Grant No. 2020750 from the United States-Israel Binational Science Foundation (BSF) and Grant No. 2109066 from the United States National Science Foundation (NSF), and by the Ministry of Science \& Technology, Israel. EZ acknowledgde funding from the Swedish National Space Agency and grant 2022-03804 from the Swedish Research Council.

\input{Builders_Acknowledgments}

We thank the CANUCS team for generously providing early access to their
proprietary data of M0416.

\end{acknowledgements}

\bibliographystyle{aa}
\bibliography{MyBiblio} 


\begin{appendix}

\begin{table*}
 \centering
  \begin{minipage}{180mm}
    \caption{Photometry of LS1 in the stacked three PEARLS epochs.}
 \label{tab_1}
 \begin{center}
 \begin{tabular}{|ccccccccccc|}   
 \hline
     \small{F435W}     &    \small{F606W}     &     \small{F814W}     &    \small{F090W}     &    \small{F115W}     &    \small{F150W}     &    \small{F200W}     &    \small{F277W}     &    \small{F356W}     &    \small{F410M}     &     \small{F444W}    \\
 \hline   
  \small{$29.9\!\pm\!1.0$} & \small{$28.8\!\pm\!0.6$} & \small{$28.5\!\pm\!0.5$} & \small{$28.3\!\pm\!0.3$} & \small{$28.1\!\pm\!0.2$} & \small{$28.1\!\pm\!0.2$} & \small{$28.0\!\pm\!0.3$} & \small{$27.7\!\pm\!0.3$} & \small{$27.8\!\pm\!0.4$} & \small{$27.7\!\pm\!0.3$} & \small{$27.6\!\pm\!0.5$} \\
\hline
\end{tabular}
 \end{center}
 \end{minipage}
\end{table*}

\section{Photometry}
\label{sec_Phot}

\begin{figure*}
\begin{center}
     \includegraphics[width=18cm]{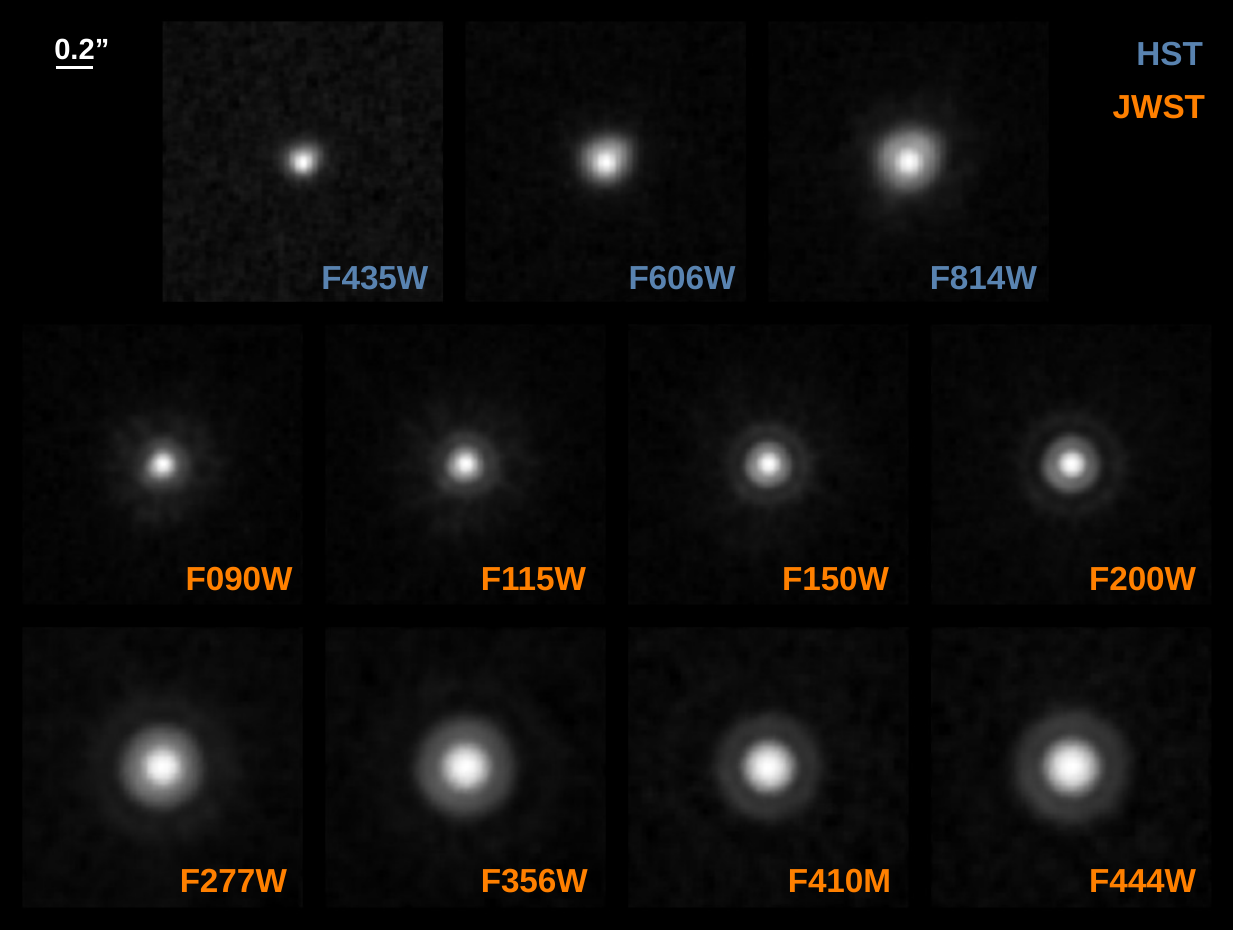}
      \caption{Stacked signal of six unsaturated stars in the same field in different bands and after coadding all three epochs (for JWST). The stars have been aligned before stacking using steps of 3 mas. In order to better show the sidelobes and diffraction spikes, we plot the logarithm of the stacked signal. The stacked signal is used as a model for the PSF in each filter.  
         }
         \label{Fig_StackedStar}
    \end{center}
\end{figure*}

\begin{figure*} 
   \includegraphics[width=18cm]{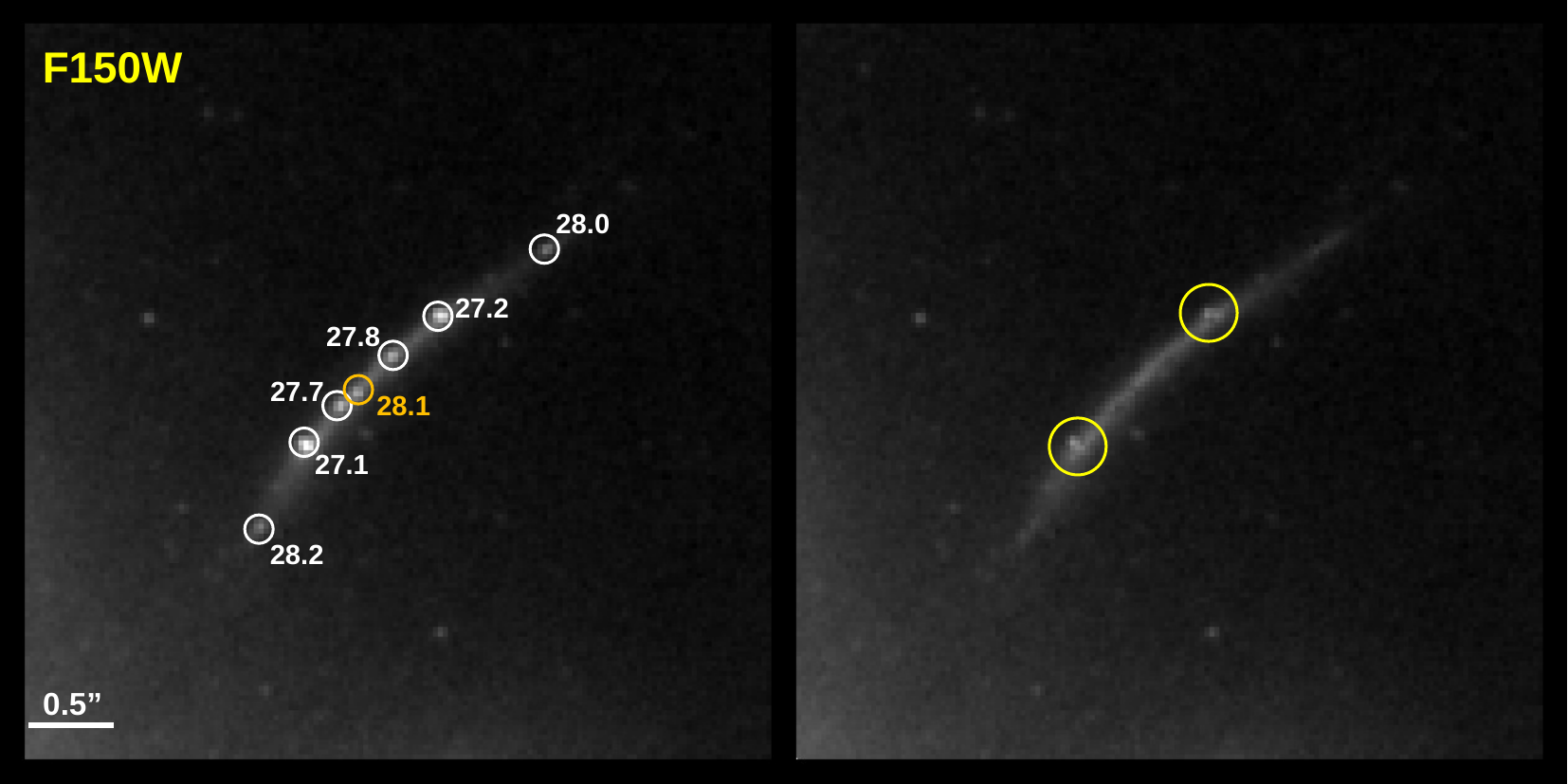}
      \caption{Example of flux estimation using the stacked star PSF model. The circles mark the position of seven sources for which the magnitudes are obtained after fitting to the stacked profile of six nearby (unsaturated) stars. The numbers show the estimated magnitude in the F150W band. The left panel shows the original data and the right panel shows the data after point source subtraction. LS1 is marked with a yellow circle in the left panel. The two yellow circles on the right panel show two regions around the brightest multiply lensed point source, b and b$'$, where additional unresolved features (also multiply lensed) can still be observed after the source removal. The same feature is observed in other filters suggesting that this is a real feature, and not an artifact. 
         }
         \label{Fig_PSfit}
\end{figure*}


\begin{figure*}
\begin{center}
     \includegraphics[width=18cm]{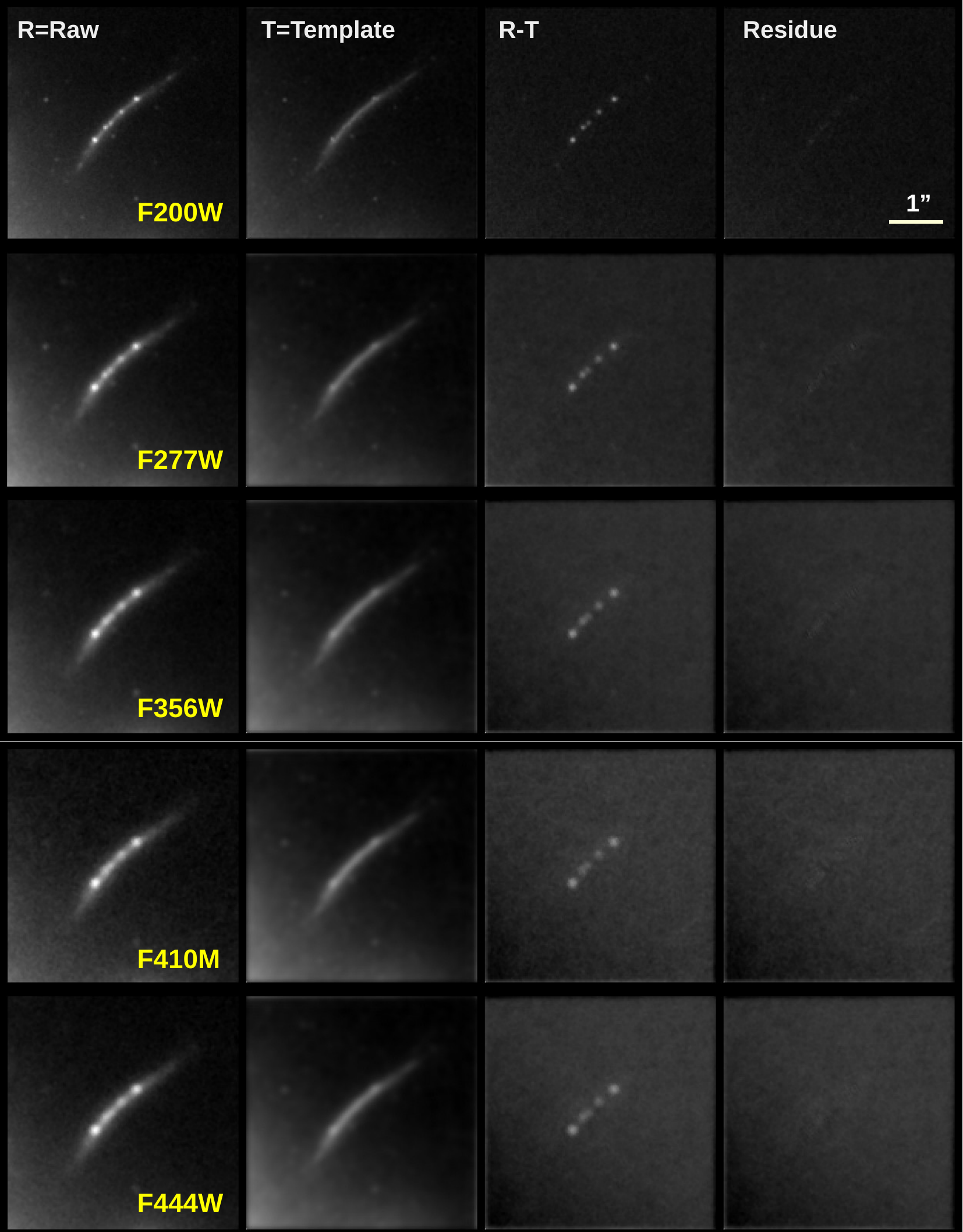}
      \caption{Illustration of the process followed to estimate the magnitudes of unresolved sources in the arc at 2 micron and larger wavelengths. In all cases the template is the point source subtracted F150W image after degrading the resolution to the one corresponding to each band.  The template is shown in the right panel of Figure~\ref{Fig_PSfit}, and at its native resolution.
         }
         \label{Fig_Clean_F200WtoF444W}
    \end{center}
\end{figure*}

To estimate the flux of the point sources in the arc, including LS1, we adopt a data-driven model for the PSF in each band. The advantage of this approach over the use of a precomputed model (such as WebPSF) is that the data-driven PSF model is self-consistent with the data. For example, the diffraction spikes of the stacked stars are exactly in the same orientation as in the targets. The constant  perforation of the mirror by microasteroids results in  degradation of the PSF, in particular its tails. Thermal variability across the mirror can result also in minute corrections to the PSF that may no be properly captured by the PSF model. Finally, since we are stacking three epochs with three different position angles, the use of stars instead of a PSF model guarantee also the correct orientation of all instrumental effects. We construct our data driven PSF model from six stars that are found near LS1. These stars are not saturated so the flux in the central region is not affected by saturation. We select a region of 50x50 pixels around each star and supersample this region with a pixel size of 3 mas. Then align the stars using a simple Gaussian model centered in the central pixel for the main peak of the star such that the star is aligned with this Gaussian model (and to  within 3 mas accuracy). After aligning the six stars to the same Gaussian model we stack their signal. The resulting average signal for the star model is shown in Figure~\ref{Fig_StackedStar} for the eleven filters considered in this work (three from HST plus eight from JWST). \\

Using this model for the PSF we estimate the flux of the seven point sources in the arc containing LS1. These seven point sources are shown in Figure~\ref{Fig_PSfit} for the case of F150W. LS1 corresponds to the central point and is highlighted with a yellow circle. The other six point sources correspond to 3 objects in the source that are multiply lensed. The flux for each source is obtained after subtracting the corresponding PSF model re-scaled by the total flux. As an example, we show the resulting image in the F150W filter, and after subtracting all seven point sources, on the right side of  Figure~\ref{Fig_PSfit}. For the two brightest sources in the arc, we find a residue near the position of these two point sources which appears repeated on both sides of the arc, suggesting that one is a counterimage of the other. These two features are marked with yellow circles in the right panel.  

The same operation is repeated for all filters. For the filters F200W, F277W, F356W, F410M and F444W we perform a more careful subtraction since it is more difficult to distinguish the point sources from the underlying arc.  For these filters we use the clean image obtained in the F150W (right panel of Figure~\ref{Fig_PSfit}) as a model, or template, for the arc. Under the assumption that the arc has similar spectral features along the entire arc, we degrade the resolution of the template to the resolution of the other filters and subtract it before performing the photometric measurement of the point sources. The degradation is done in Fourier space by re-scaling each Fourier mode with wavenumber $k$ by the factor $W(k)= \sqrt{P_R(k)/P_T(k)}$, where $P_R(k)$ is the power spectrum of the raw data from which we want to subtract the arc, and $P_T(k)$ is the power spectrum of the template (in our case the F150W data after point source subtraction). Once the template has been degraded to the desired resolution, we subtract it from the raw data (at the same resolution), and we estimate the flux of the remaining point sources. This process is graphically shown in Figure\ref{Fig_Clean_F200WtoF444W}. The left column shows the raw data in each filter. The second column shows the template degraded to the resolution of the raw data following the process described above. The third column shows the difference Raw-Template. In this difference, the point sources are evident in all bands, while some are not so clearly visible in the raw data. Finally, the last column shows the residue left after fitting the point sources with the corresponding star shapes from Figure~\ref{Fig_StackedStar}. The residue has very little signal left, specially in the longest wavelengths. 
The final photometry for LS1 and for the stacked three epochs of the PEARLS program is shown in table~\ref{tab_1}

\section{Band differences}
\label{sec_CornerPlot}

\begin{figure*}
\begin{center}
     \includegraphics[width=18cm]{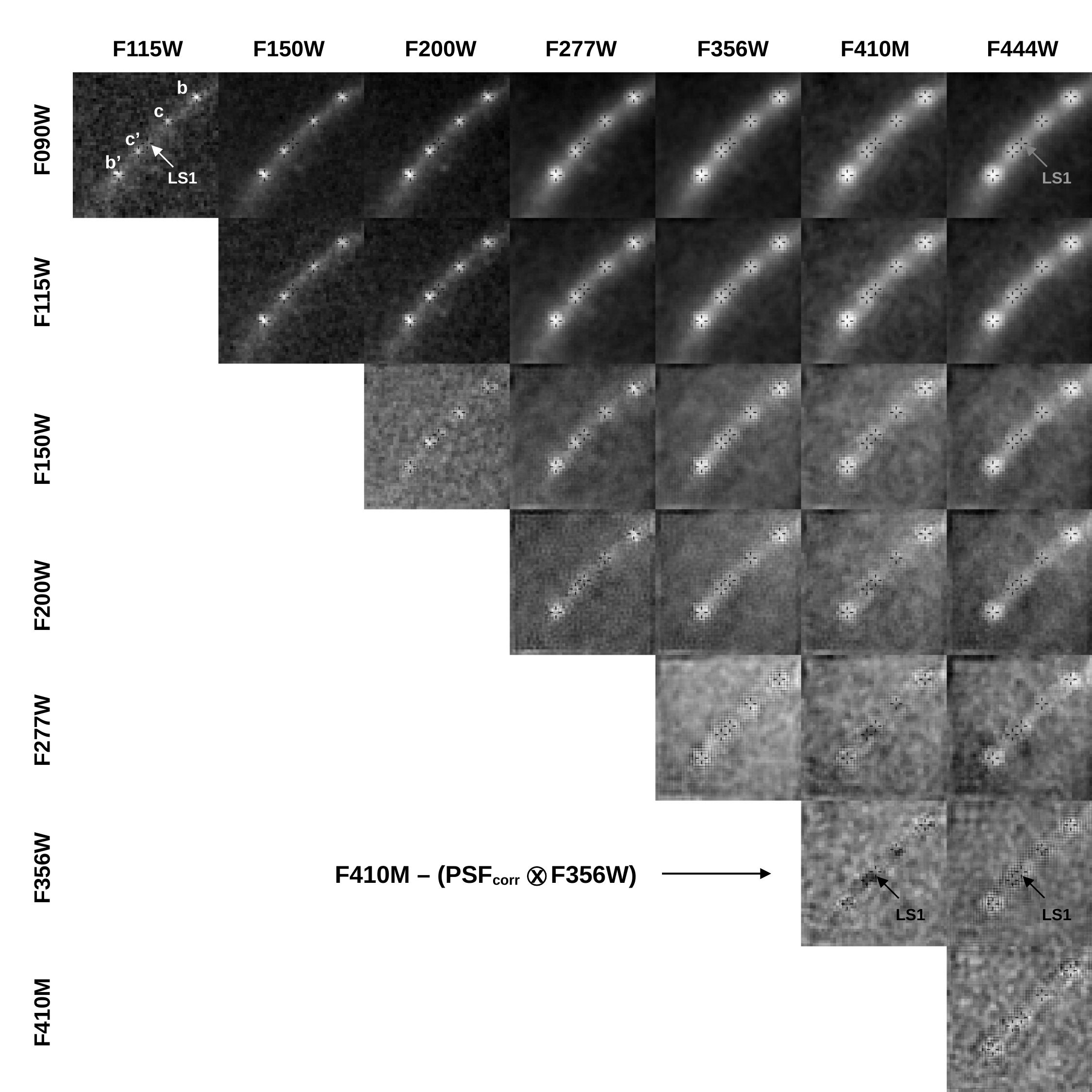}
      \caption{Simple image subtraction M1 with all NIRCam combinations. The positive image is shown by labels above each column, and the negative image is shown by labels to the left of each row.  For example, the top-left panel shows $\rm F115W-F090W$. Sources are labeled in the top-left panel with an arrow showing the location of LS1.  Crosses in all panels show the source positions. 
         }
         \label{Fig_CornerPlot1}
    \end{center}
\end{figure*}

\begin{figure*}
\begin{center}
     \includegraphics[width=18cm]{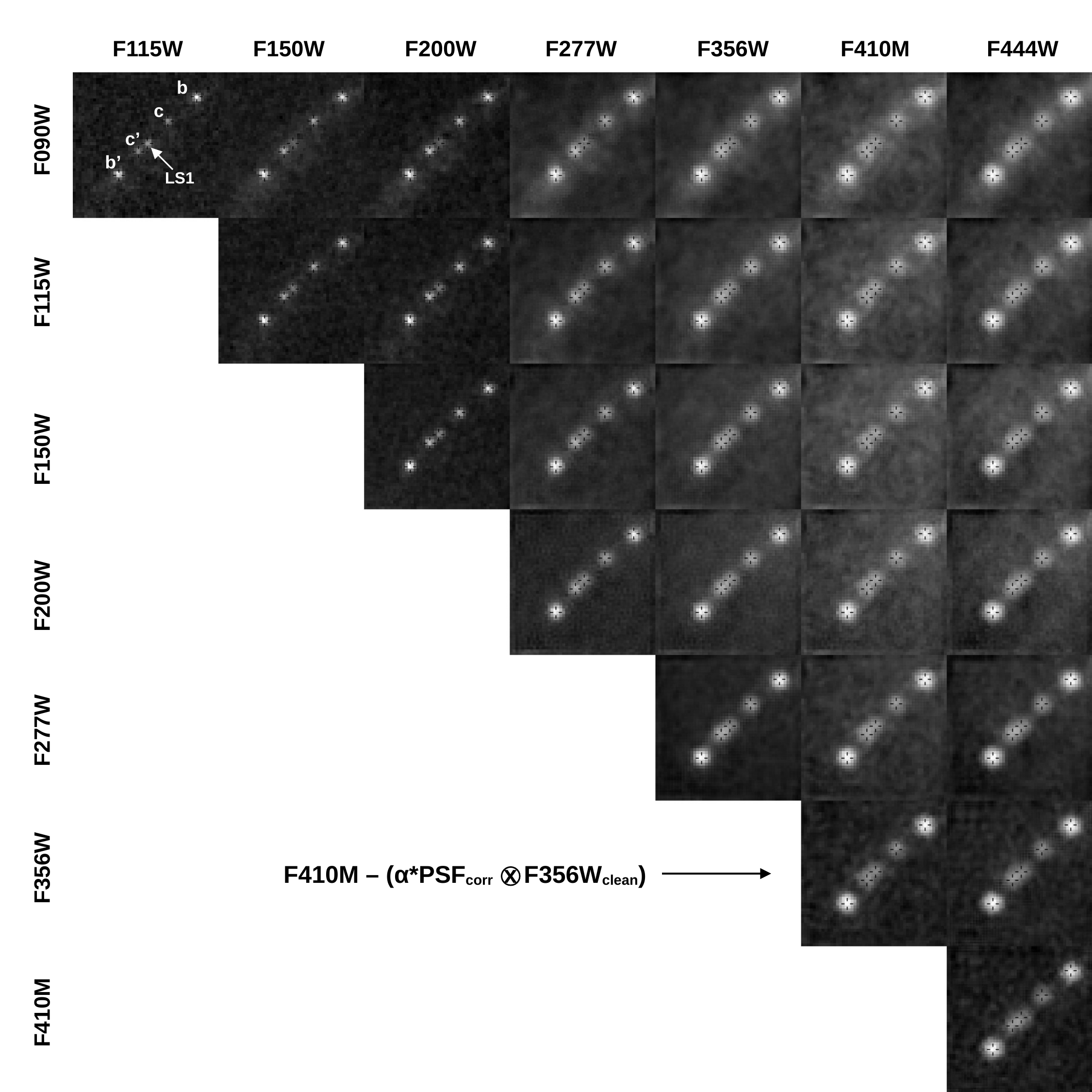}
      \caption{Image subtraction M2 removing the arc. Other details are the same as Figure~\ref{Fig_CornerPlot1}.
         }
         \label{Fig_CornerPlot2}
    \end{center}
\end{figure*}

\begin{figure*}
\begin{center}
     \includegraphics[width=18cm]{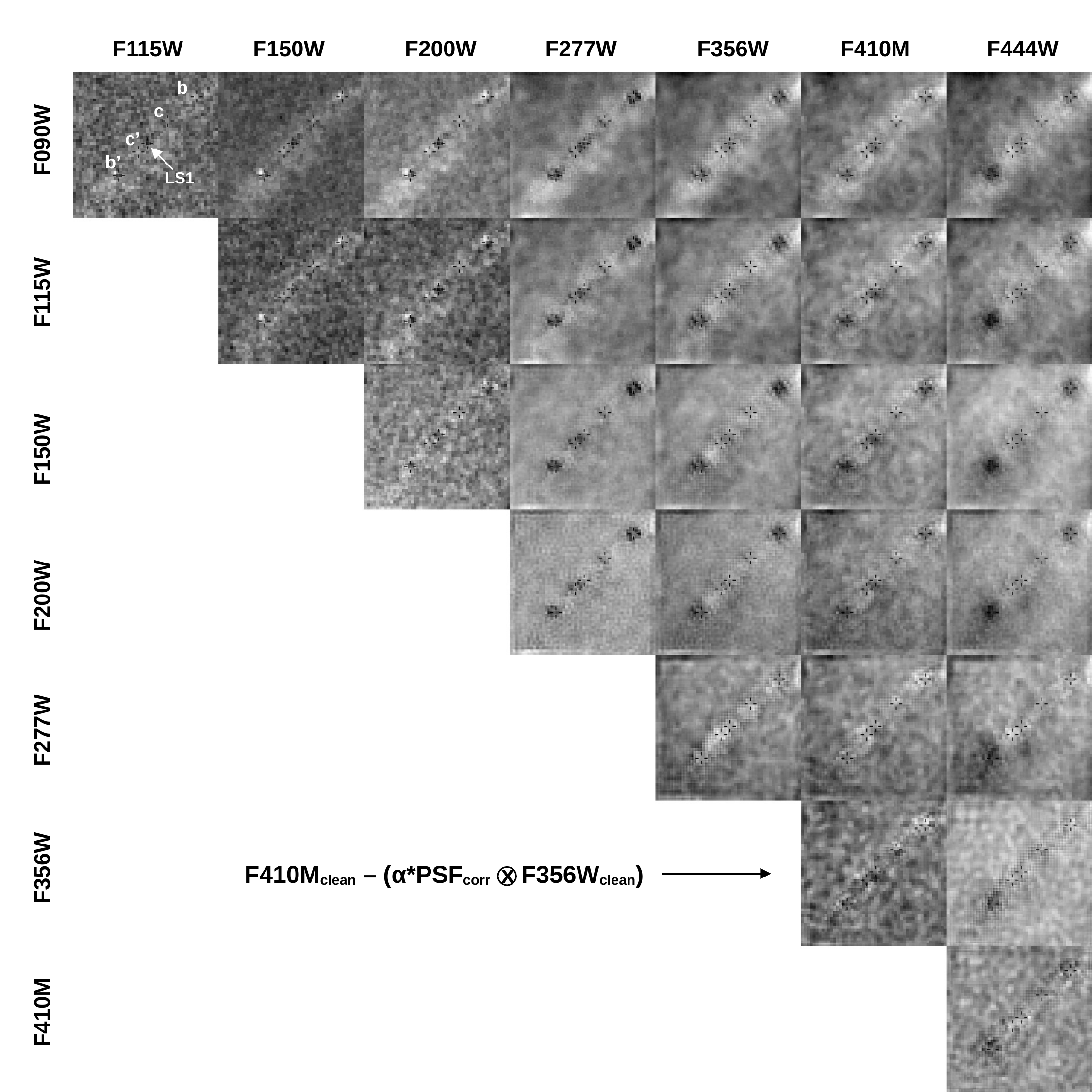}
      \caption{Image subtraction M3 removing the arc and the point sources. Other details are the same as Figure~\ref{Fig_CornerPlot1}.   
         }
         \label{Fig_CornerPlot3}
    \end{center}
\end{figure*}

Image differences can enhance faint details compared to single-band images, especially when the details are hidden by brighter nearby features.  To search for a possible LS1 counterimage or anything else hidden in the arc, made difference images from each pair of NIRCam images. In all cases, the shorter-wavelength, higher-resolution image was first convolved with a Gaussian kernel to match the longer-wavelength, lower-resolution image in the pair, then subtracted.  Three distinct methods were used:
\begin{enumerate}
\item[M1:] a simple image subtraction after matching the resolution.  Results are shown in
Figure~\ref{Fig_CornerPlot1}. In this scheme, any source with a flat SED will show zero flux.  LS1 is very faint in all images because it has a nearly flat SED, but the arc is seen because its  SED rises with wavelength.
\item[M2:] an attempt to subtract only the arc.
The seven unresolved sources (a, b, c, LS1, c$'$, b$'$, and a$'$) were first subtracted from the shorter-wavelength image, that image was then normalized by a constant factor $\alpha$ and convolved to the resolution of the longer-wavelength image, and that image was subtracted.  The factor $\alpha$ was chosen for each image pair to minimize the residual of the arc.
Figure~\ref{Fig_CornerPlot2} shows the result.
The arc disappears, as expected, but LS1 shows up  well.  These images were used for photometry of LS1 (Section~\ref{sec_SED}.  A faint negative image just northwest of LS1 can be seen in the F200W column indicating the possible presence of a source bluer than the arc.
\item[M3:] an attempt to subtract both point sources and the arc.  The seven unresolved sources were subtracted from {\em both} images, then the shorter-wavelength image was convolved,  normalized, and subtracted.
Figure~\ref{Fig_CornerPlot3} shows the results.  Residuals are noticeable for sources b and b$'$ because they are not well described by a single point source. The faint negative image again shows up in the F200W column and in $\rm F150W-F090W$. LS1 has a very small residual in all images, showing that the photometry in Figure~\ref{Fig_SED_LS1} is accurate.
\end{enumerate}

\section{Profile along the arc}
\begin{figure*}
\begin{center}
     \includegraphics[width=8cm]{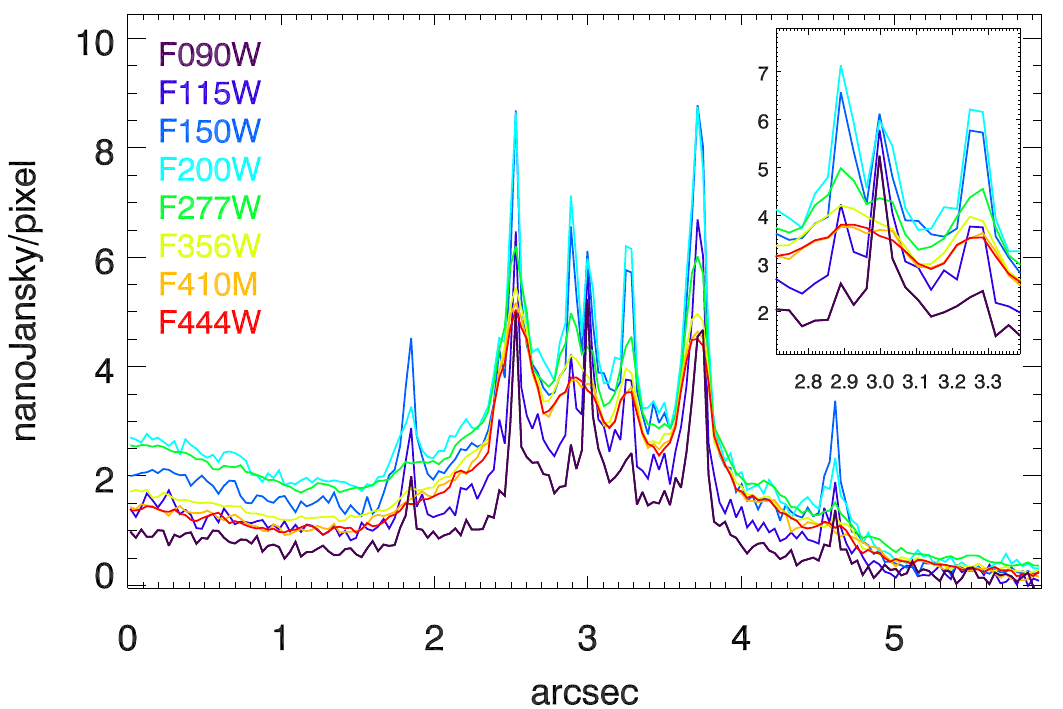}
     \includegraphics[width=8cm]{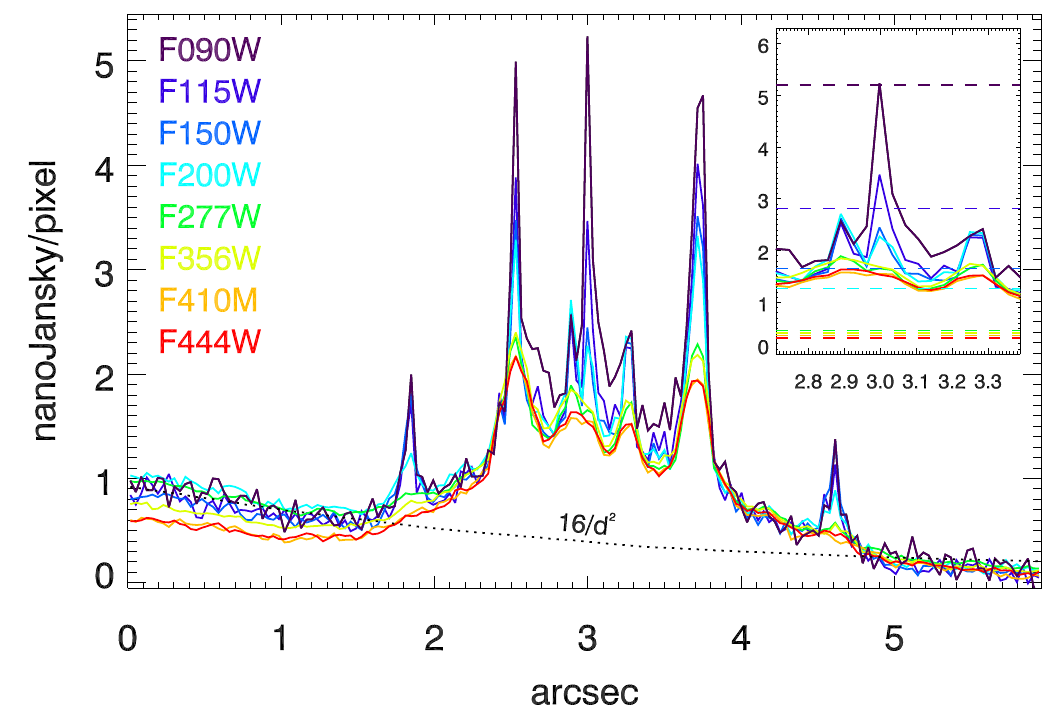}
      \caption{Profile along the arc. The arc hosting Mothra can be very well described as a portion of a circle with radius 3.795" and centered in RA=64.0359850, DEC=-24.0669947. The figure on the left shows the one dimensional profile along this arc. Mothra is the peak in the middle at 3 arcsec. The inset plots zooms-in around the position of Mothra. The figure on the right is similar but all profiles have been re-scaled to the amplitude of the F090 profile at $\approx 4.2$ arcsec. The black doted line is a power law model that traces the ICL. Here $d$ is the distance to the BCG expressed in arcseconds. For reference, Mothra is at 6.3" from the BCG. 
      The horizontal dashed lines in the inset plot show the combined effect of the PSF dilution and re-scaling. A source with the same flux in all bands as in F090W that is convolved by the PSF of each band  would have its amplitude at the corresponding dashed lines.
         }
         \label{Fig_ProfCirc1}
    \end{center}
\end{figure*}

In this appendix we present the one dimensional profile of the arc hosting Mothra and along the arc. To do this we find the circle that best fits the arc. We find that a circle with radius 3.795" and centered in RA=64.0359850, DEC=-24.0669947 passes through all seven point sources identified in the arc. The one-dimensional profile for all JWST bands is shown in Figure~\ref{Fig_ProfCirc1}. The left panel shows the profiles of the raw data, while in the right panel we re-scale all profiles by a multiplicative bias factor in order to bring them to the same amplitude of the F090W profile at $\approx 4.2"$.  In each figure we include a zoom-in around the Mothra peak. The profiles are contaminated by the emission of the ICL which varies with frequency. At 0", the flux is dominated by the ICL which peaks at 2 micron. The figure on the right has all amplitudes re-scaled to the amplitude of the F090W profile at $\approx 0.42"$. This is done to reduce the impact of the host galaxy and ICL and better show the relative brightness of the peaks. Since each band has a different resolution, and the amplitude of a point source is reduced with poorer resolution, we show as dashed lines the expected change in flux for a source with constant flux (and equal to the flux of the source in F090).

\section{Offsets from VSVPA}\label{sec_VSVPA}
When studying the time variability of Mothra in section~\ref{sec_timevar}, we discuss at the end of that section a small offset in the position of the difference of epochs. 
In this appendix we study these offsets using mock data. \\ 
We identify an effect which we call Varying Source Varying Position Angle (VSVPA), that manifests itself in the difference between epochs when these are done with different position angles and at the position of sources that are varying in flux. The combined effect of varying flux and varying PSF results in residuals in the difference between epochs that are not necessarily centered with the position of the source. This apparent offsets are due to asymmetries in the PSF that are more evident when the flux is changing between epochs. To test this hypothesis we create a simulation that mimics the real observation around the arc at z=2.097. 

\begin{figure*}
\begin{center}
     \includegraphics[width=16cm]{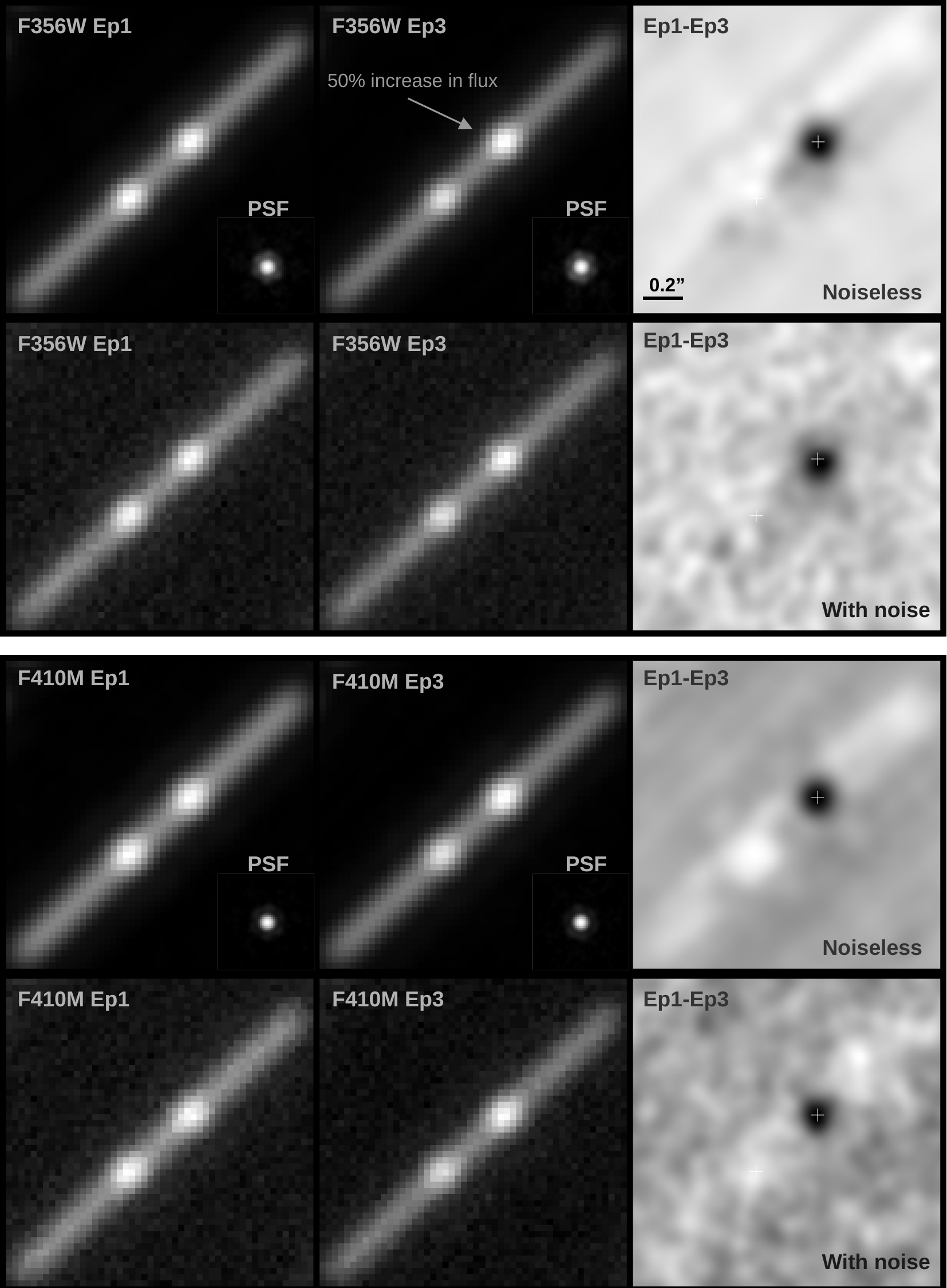}
      \caption{Illustrations of the VSVPA effect with mock data. The top two panels show the mock data for epochs 1 and 3 and their difference for the band F356W, with and without instrumental noise. The bottom two panels are similar but for the F410M band. 
      Offsets of approximately 0.07" can be appreciated in the difference between epochs. These offsets can be appreciated in some filters, such as F356W, even in the absence of noise.  In all cases the images are 1.5" across. The white cross marks the position of the varying point source. The length of the arms of the cross is 0.06" (end-to-end).
         }
         \label{Fig_VSVPA}
    \end{center}
\end{figure*}

 First we simulate a thin arc at the same orientation as our arc (42.8 degrees counterclockwise from west). Then we add two point sources along the arc with approximately the same flux ratio between Mothra and the underlying arc. Then we convolve with the PSF from epoch 1 in different filters. For epoch 3 we do the same but before convolving by the PSF of epoch 3 we increase the flux of one of the point sources by 50\%, which is approximately the estimated relative increase in flux of Mothra between epochs 1 and 3. Then we add Gaussian noise to each epoch with similar standard deviation to the one from the real data. Finally we subtract epoch 3 from epoch 1. The different steps and final result are shown in Figure~\ref{Fig_VSVPA} for two filters, F356W (first two rows), for which we observe an offset in JWST data of $\approx 0.07"$ between the position of Mothra and the position of the excess flux, and F410M (bottom two rows) for which we do not observe an offset when doing the difference between epochs. In both cases we show the results with and without noise, to better show the effect due to the different PSFs.  In each row, Epoch1 is on the left and Epoch3 in the middle. The difference Epoch1-Epoch3 is shown on the right panel. This difference is smoothed with a  0.09" Gaussian as we did with the real data.

\begin{figure}
\begin{center}
     \includegraphics[width=16cm]{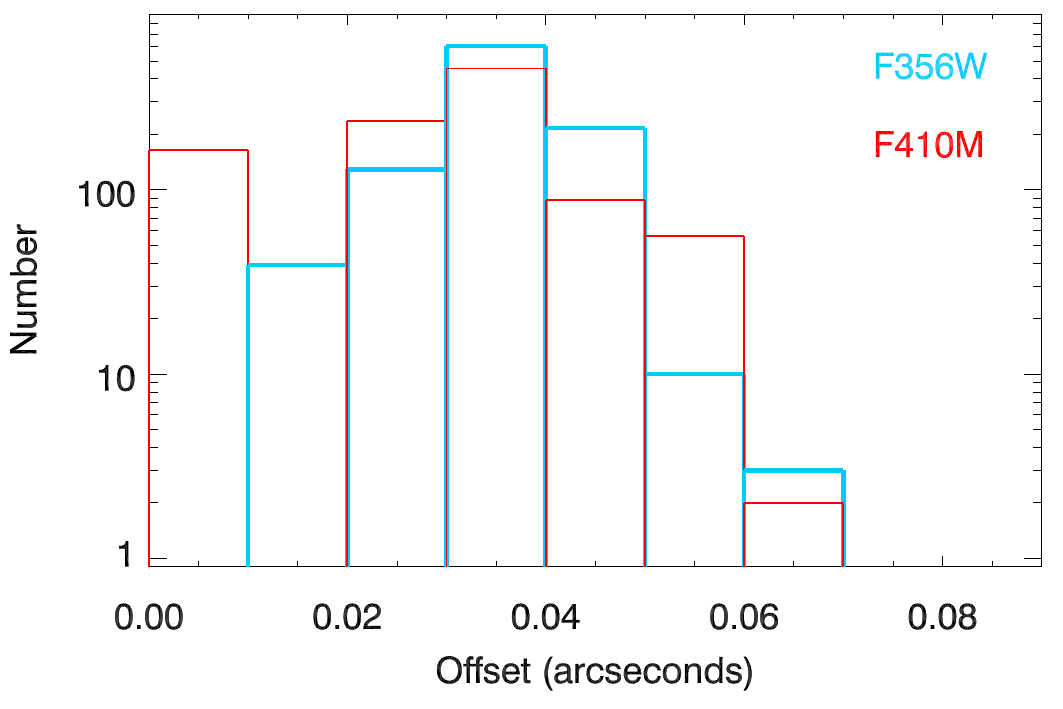}
      \caption{Distribution of offsets from 1000 realizations.
         }
         \label{Fig_Offsets}
    \end{center}
\end{figure}

We repeat this experiment 1000 times, each time with a different realization fo the noise. The measured offsets are shown in Figure~\ref{Fig_Offsets}. 
An offset (comparable in magnitude to the one observed in the real data) is appreciated in a significant fraction of these realizations. Sources that are not varying in time, such as the point source in the southern portion of the arc, do not show up in the residual. Hence, in order to see this effect a varying source is also needed. The effect can then be explained as a combination of a varying source and a varying PSF (its position angle). We have not conducted an exhaustive study of this effect (since it is well beyond the scope of this paper), but we expect it to depend on the position angle and choice of filters. Also, we notice that for brighter sources that have larger variations in flux the offset is smaller. We conclude then that the offsets observed in the data can be satisfactorily explained as an instrumental effect combined with an intrinsically varying source. 

\end{appendix}

\end{document}

%% file: Builders_Acknowledgments.tex
This work is based on observations made with the NASA/ESA/CSA James Webb Space
Telescope. The data were obtained from the Mikulski Archive for Space
Telescopes at the Space Telescope Science Institute, which is operated by the
Association of Universities for Research in Astronomy, Inc., under NASA
contract NAS 5-03127 for JWST. These observations are associated with JWST
programs 1176 and 2738.
RAW, SHC, and RAJ acknowledge support from NASA JWST Interdisciplinary
Scientist grants NAG5-12460, NNX14AN10G and 80NSSC18K0200 from GSFC. Work by
CJC acknowledges support from the European Research Council (ERC) Advanced
Investigator Grant EPOCHS (788113). BLF thanks the Berkeley Center for
Theoretical Physics for their hospitality during the writing of this paper.
MAM acknowledges the support of a National Research Council of Canada Plaskett
Fellowship, and the Australian Research Council Centre of Excellence for All
Sky Astrophysics in 3 Dimensions (ASTRO 3D), through project number CE17010001.
CNAW acknowledges funding from the JWST/NIRCam contract NASS-0215 to the
University of Arizona.
We also acknowledge the indigenous peoples of Arizona, including the Akimel
O'odham (Pima) and Pee Posh (Maricopa) Indian Communities, whose care and
keeping of the land has enabled us to be at ASU's Tempe campus in the Salt
River Valley, where much of our work was conducted.